\documentclass[twocolumn,twocolappendix]{aastex631}
\usepackage{makecell}
\usepackage {verbatim}
\usepackage[utf8]{inputenc}
\usepackage{amsmath, amsfonts, amssymb, graphics,graphicx, wrapfig, nicefrac}
\usepackage{epsfig}
\usepackage{epstopdf}
\usepackage{multirow}
\usepackage{graphicx}
\usepackage{rotating}
\usepackage{ulem}
\usepackage{booktabs}

\newcommand{\hmole}{H$_2$}

\newcommand{\htcs}{H$_{2}$CS}
\newcommand{\ttso}{$^{33}$SO}
\newcommand{\sot}{SO$_{2}$}
\newcommand{\tfsot}{$^{34}$SO$_{2}$}
\newcommand{\ttsot}{$^{33}$SO$_{2}$}
\newcommand{\octfs}{OC$^{34}$S}
\newcommand{\seoo}{SO$^{18}$O}
\newcommand{\tone}{$T(\rm H_{2}CS)$}
\newcommand{\ttwo}{$T(\rm ^{34}SO_{2})$}

\shorttitle{Sulfur-bearing molecules in dense cores}
\shortauthors{Tang et al.}
\begin{document}

\title{A survey of sulfur-bearing molecular lines toward the dense cores in eleven massive protoclusters}

\correspondingauthor{Mengyao Tang}
\email{mengyao\_tang@yeah.net}

\author[0000-0001-9160-2944]{Mengyao Tang}
\affiliation{Institute of Astrophysics, School of Physics and Electronic Science, Chuxiong Normal University, Chuxiong 675000, PR China}

\author[0000-0003-2302-0613]{Sheng-Li Qin}
\affiliation{School of Physics and Astronomy, Yunnan University, Kunming, 650091, PR China}

\author[0000-0002-5286-2564]{Tie Liu}
\affiliation{Shanghai Astronomical Observatory, Chinese Academy of Sciences, Shanghai 200030, PR China}

\author[0000-0003-2343-7937]{Luis A. Zapata}
\affiliation{Instituto de Radioastronom\'ia y Astrof\'isica, Universidad Nacional Aut\'onoma de M\'exico, P.O. Box 3-72, 58090, Morelia, Michoac\'an, M\'exico}

\author[0000-0001-8315-4248]{Xunchuan Liu}
\affiliation{Shanghai Astronomical Observatory, Chinese Academy of Sciences, Shanghai 200030, PR China}

\author[0000-0001-5703-1420]{Yaping Peng}
\affiliation{Department of Physics, Faculty of Science, Kunming University of Science and Technology, Kunming 650500, PR China}

\author[0000-0001-5950-1932]{Fengwei Xu}
\affiliation{Kavli Institute for Astronomy and Astrophysics, Peking University, Beijing 100871, PR China} \affiliation{Department of Astronomy, School of Physics, Peking University, Beijing, 100871, PR China}

\author[0000-0002-5682-2906]{Chao Zhang}
\affiliation{Institute of Astronomy and Astrophysics, School of Mathematics and Physics, Anqing Normal University, Anqing, PR China}

\author[0000-0002-8149-8546]{Ken'ichi Tatematsu}
\affiliation{Nobeyama Radio Observatory, National Astronomical Observatory of Japan, National Institutes of Natural Sciences, 462-2 Nobeyama, Minamimaki, Minamisaku,
	Nagano 384-1305, Japan}
\affiliation{Department of Astronomical Science, The Graduate University for Advanced Studies, SOKENDAI, 2-21-1 Osawa, Mitaka, Tokyo 181-8588, Japan}



\begin{abstract}
Sulfur-bearing molecules are commonly detected in dense cores within star-forming regions, but the total sulfur budget is significantly low, when compared to the interstellar medium (ISM) value.
The properties of sulfur-bearing molecules are not well understood due to the absence of large sample studies with uniform observational configurations.
To deepen our understanding of this subject, we conducted a study using ALMA 870 \micron~observations of 11 massive protoclusters. 
By checking the spectra of 248 dense cores in 11 massive protoclusters, a total of 10 sulfur-bearing species (CS, SO, \htcs, NS, \sot, \ttso, \tfsot, \ttsot, \seoo, \octfs) were identified.
The parameters including systemic velocities, line widths, gas temperatures, column densities, and abundances were derived. Our results indicate that SO appears to be more easily detected in a wider range of physical environments than \htcs, despite these two species show similarities in gas distributions and abundances. \tfsot~and \htcs~are good tracers of the temperature of sulfur-bearing species, in which \htcs~traces the outer warm envelope and \tfsot~is associated with high-temperature central-regions. High-mass star-forming feedback (outflow and other non-thermal motions) significantly elevates the sulfur-bearing molecular abundances and detection rates specifically for \sot~and SO. A positive correlation between the \sot~abundance increasing factor ($F$) and temperatures suggests that \sot~could serve as a sulfur reservoir on the grain mantles of dense cores and then can be desorbed from dust to gas phase as the temperature rises. This work shows the importance of a large and unbiased survey to understand the sulfur depletion in dense cores.

\end{abstract}


\section{Introduction} \label{sec:introduction}
Sulfur is the tenth most abundant element in our galaxy, and sulfur-bearing (S-bearing, hereafter) molecules are widely detected in different environments of star formation, e.g., circumstellar envelopes \cite[]{van2003A&A...412..133V,Zapata2019ApJ...872..176Z}, prestellar core \cite[]{Vastel2018MNRAS.478.5514V}, hot core \cite[]{Charnley1997ApJ...481..396C,Hatchell1998A&A...338..713H,Wilson1994ARA&A..32..191W,Vidal2018MNRAS.474.5575V}, and shock regions \cite[]{Bachiller1997ApJ...487L..93B,Bachiller2001A&A...372..899B,Wakelam2005A&A...437..149W}.
Consequently, S-bearing molecules have attracted broad attention and have generally been adopted as molecular probes to investigate the properties of star-forming regions \cite[]{Artur2019A&A...626A..71A,Oya2019ApJ...881..112O,Tychoniec2021A&A...655A..65T,Artur2023A&A...678A.124A}.

However, the properties of S-bearing species are still unclear, a persistent unsolved problem in the research on sulfur is so called sulfur-depletion problem.
Since the 1970s, the sulfur depletion problem became evident because it was found that the observed abundances of S-bearing molecules in dense regions generally lower than diffused regions (S/H $\sim$10$^{-5}$) \cite[]{Penzias1971ApJ...168L..53P,Oppenheimer1974ApJ...187..231O,Tieftrunk1994A&A...289..579T,Palumbo1995ApJ...449..674P}.
Since then, sulfur depletion has always been a mystery in the research field of astronomy.
\cite{Ruffle1999MNRAS.306..691R} suggested that sulfur predominantly existed as S$^{+}$ in translucent gas, frozen quickly onto negatively charged grains during cloud collapse.
Chemical models suggest that hydrogenation is most effective on grains, and H$_{2}$S is a main sulfur reservoir in grain mantles \cite[][]{Charnley1997ApJ...481..396C}.
However, H$_{2}$S has not been identified in the solid phase, with its gas phase abundance estimated to be $\sim$10$^{-7}$ \cite[e.g.][]{Minh1990ApJ...360..136M,vanDi1998ARA&A..36..317V}, notably lower than sulfur levels in diffuse regions. 
Over the last two decades, various potential candidates for solid sulfur reservoirs have suggested, including hydrated sulphuric acid (H$_{2}$SO$_{4}$$\cdot$$n$H$_{2}$O), H$_{2}$S$_{n}$, sulfide minerals (FeS and MgS), sulfur chains (S$_{n}$), organo-sulfur species \cite[][]{Keller2002Natur.417..148K,Hony2002A&A...393L.103H,Scappini2003MNRAS.341..657S,Wakelam2004A&A...422..159W,Jim2011A&A...536A..91J,Druard2012MNRAS.426..354D,Vidal2017MNRAS.469..435V,Laas2019A&A...624A.108L,Kama2019ApJ...885..114K}.
Regrettably, detection in interstellar ices has only been successful for OCS \cite[]{Geballe1985A&A...146L...6G,Palumbo1995ApJ...449..674P,McClure2023NatAs...7..431M} and \sot~\cite[]{Boogert1997A&A...317..929B,Yang2022ApJ...941L..13Y,McClure2023NatAs...7..431M, Rocha2024A&A...683A.124R}, leaving sulfur depletion as an ongoing enigma.
	
The bulk of the sulfur reservoir is proposed to be locked onto grain mantels during the collapsing process \cite[][]{Ruffle1999MNRAS.306..691R,Woods2015MNRAS.450.1256W,Vidal2017MNRAS.469..435V,Navarro2020A&A...637A..39N}, and is subsequently released into gas phase when the temperature exceeds 100 K. Therefore, the collapsing regions could be ideal for investigating sulfur depletion.
The first aim of this study is to investigate the sulfur depletion problem in massive star-forming regions. 
Therefore, we initiated a study using the data of an ALMA survey of 11 massive protoclusters in band 7.
These 11 protoclusters, situated at distances ranging from 2.6 to 7.6 kpc, were recognized as infalling clumps with``blue-profile'' signatures \cite[][]{Liu2016ApJ...829...59L,Yue2021RAA....21...14Y}.
Moreover, observations of other star-forming activities such as outflows \cite[][]{Baug2021MNRAS.507.4316B} and H$_{\rm II}$ regions \cite[][]{Xu2024ApJS..270....9X} suggest that these clumps are active star-forming regions. 

The second main purpose of this study is to explore the similarities and differences among properties of S-bearing species in various dense cores. Despite numerous works conducted in past decades, most are performed by single-dish telescopes or from different observation projects with different observational frequency ranges, angular resolutions, and sensitivities. 
Recent observations of C$^{34}$S and \sot~ in 10 sources
\cite[][]{Artur2019A&A...626A..71A} and CS, SO, \sot, and $^{34}$SO in 50 sources \cite[][]{Artur2023A&A...678A.124A} have revealed interesting statistical findings, suggesting that conducting statistical analyses on larger sample sizes could yield more valuable insights. 
However, their studies focused on low-mass star-forming regions like Ophiucus and Perseus with a limited number of detected species and samples. The lack of statistical results from extensive samples of dense cores in star-forming regions, particularly high-mass star-forming regions, poses a significant obstacle to advancing research on S-bearing species. Therefore, a statistical investigation involving several hundred dense cores spanning diverse physical environments and evolutionary stages could greatly enhance  comprehension of S-bearing species.
	
In our seuveyed 11 protoclusters, a total of 248 dense cores were identified before \cite[][]{Xu2024ApJS..270....9X}. 
These 248 dense cores cover diverse physical environments, including high-mass, low-mass, prestellar, protostellar, with/without outflows, and hot core/corinos \cite[][]{Qin2022MNRAS.511.3463Q,Xu2024ApJS..270....9X,Liu2023ApJ...958..174L}. They serve as ideal subjects for statistically exploring the characteristics of S-bearing species. A comparative analysis of S-bearing species' parameters across various environments can unveil both similarities and differences among these species, advancing our knowledge of sulfur chemistry.

The paper is structured as follows: Section~\ref{sec:observation} introduces the observational information and the samples of dense cores. 
Section~\ref{sec:results} presents detected species, spectra, and derived parameters. 
In Section~\ref{sec:discussion}, statistical analyses and sulfur depletion problem are discussed.
Section~\ref{sec:summary} summarizes the key highlights of the paper. 
The appendix contains additional information related to our results.

\section{Observations} \label{sec:observation}
\subsection{870 \micron~data}
\begin{deluxetable*}{cccccccc} 
	\label{tab:observation}
	\tabletypesize{\tiny}
	\tablecaption{Information of observations}
	\tablewidth{0pt} \tablehead{
		\colhead{Source name} &\colhead{Phase center} &$T_{\rm dust}$$^{a}$ &Distance$^{b}$ &\colhead{RMS noise}           &\colhead{Beam size} &\colhead{Phase calibrator}  &\colhead{Bandpass \& Flux calibrator}  \\
		\colhead{}            &\colhead{(R.A., Decl)} &(K)  &(kpc) &\colhead{(mJy beam$^{-1}$)}   &\colhead{($^{\prime\prime} \times ^{\prime\prime}$)}   &                                      
	}\startdata
	IRAS 14382-6017   & 14:42:02.76, -60:30:35.1  &28.0 &4.1 & 1.0 &1.175$\arcsec$$\times$0.793$\arcsec$  & J1524-5903 &J1427-4206 \\
	IRAS 14498-5856   & 14:53:42.81, -59:08:56.5 &26.7 &3.2 & 1.0 &1.065$\arcsec$$\times$0.785$\arcsec$  & J1524-5903 &J1427-4206 \\
	IRAS 15520-5234   & 15:55:48.84, -52:43:06.2 &32.2 &2.6 & 1.5 &0.852$\arcsec$$\times$0.699$\arcsec$  & J1650-5044 &J1924-2914,J1427-4206,J1517-2422 \\
	IRAS 15596-5301   & 16:03:32.29, -53:09:28.1  &28.5 &4.4 & 0.5 &0.824$\arcsec$$\times$0.691$\arcsec$  & J1650-5044 &J1924-2914,J1427-4206,J1517-2422 \\
	IRAS 16060-5146   & 16:09:52.85, -51:54:54.7  &32.2 &5.2 & 2.5 &0.820$\arcsec$$\times$0.681$\arcsec$  & J1650-5044 &J1924-2914,J1427-4206,J1517-2422 \\
	IRAS 16071-5142   & 16:11:00.01, -51:50:21.6     &23.9 &4.9 & 1.3 &0.810$\arcsec$$\times$0.666$\arcsec$  & J1650-5044 &J1924-2914,J1427-4206,J1517-2422 \\
	IRAS 16076-5134   & 16:11:27.12, -51:41:56.9     &30.1 &5.0 & 0.6 &0.734$\arcsec$$\times$0.628$\arcsec$  & J1650-5044 &J1924-2914,J1427-4206,J1517-2422 \\
	IRAS 16272-4837   & 16:30:59.08, -48:43:53.3 &23.1 &3.2 & 1.4 &0.799$\arcsec$$\times$0.662$\arcsec$  & J1650-5044 &J1924-2914,J1427-4206,J1517-2422 \\
	IRAS 16351-4722   & 16:38:50.98, -47:27:57.8  &30.4 &2.9 & 1.2 &0.771$\arcsec$$\times$0.651$\arcsec$  & J1650-5044 &J1924-2914,J1427-4206,J1517-2422 \\
	IRAS 17204-3636   & 17:23:50.32, -36:38:58.1  &25.8 &2.9 & 0.5 &0.790$\arcsec$$\times$0.660$\arcsec$  & J1733-3722 &J1924-2914 \\
	IRAS 17220-3609   & 17:25:24.99, -36:12:45.1   &25.4 &7.6 & 1.7 &0.813$\arcsec$$\times$0.653$\arcsec$  & J1733-3722 &J1924-2914 \\
	\enddata
	\tablecomments{
		$^{a}$: Dust temperatures listed in the table come from \citep{Urquhart2018MNRAS.473.1059U}. 
		$^{b}$: Distance of each clump is estimated by \cite{Xu2024ApJS..270....9X}. }
	\end{deluxetable*}

The data utilized in this study are from ALMA Cycle 5 observations (PI: Tie Liu, Project ID: 2017.1.00545S), conducted between 18th and 20th May 2018 (UTC).
The antenna configuration of C43-1 was adopted, employing a total of 48 12-meter antennas for the observations.
In the project, a total of 11 massive protoclusters were observed. 
The basic information of these 11 protoclusters is listed in Table \ref{tab:observation}.
To cover the entire regions of these sources, a mosaic observation mode was used to extend the field of view.
Phase, amplitude, and bandpass calibration were conducted, the calibrators used are listed in Table \ref{tab:observation}.
The flux calibration error of our observations is $\sim$10\%, which is consistent with the flux calibration accuracy in Band 7 as stated in ALMA documentation \cite[][]{almaguide}.
For the continuum data of the 11 sources, three rounds of phase self-calibration and one round of amplitude self-calibration were performed using \texttt{CASA} \cite[][]{2022PASP..134k4501C}.
In the 3 rounds of phase-only self-calibration, 
the solution intervals start with a single scan length and terminate with an integration times of 6.05s.
For the amplitude self-calibration, the solution interval ranges were set as infinite. The detailed information of self-calibration is presented in Appendix \ref{Sec:selfcal}.

During the imaging process, the deconvolution was set as ``\texttt{hogbom}'' with a weighting parameter of ``\texttt{briggs}'', and the robust parameter of 0.5 was used to balance the sensitivity and resolution of the final images.
A primary beam correction was applied.
The resulting field of view is approximately 46 arcseconds with an angular resolution of approximately 0.8 arcseconds, corresponding to spatial scales ranging from  0.01 to 0.04 pc. The largest angular scales (LAS) of the observations are from 7.2$^{\prime\prime}$ to 9.2$^{\prime\prime}$.

Four spectral windows, namely spw25, spw27, spw29, and spw31, were set to observe spectral line data. 
The center frequencies of spw25 and spw27 are 354.4 GHz and 356.7 GHz, respectively, with a spectral resolution of 0.244 MHz.
The center frequencies of spw29 and spw31 are 345.1 GHz and 343.2 GHz, respectively, with a spectral resolution of 0.976 MHz.
Each spectral window has a total of 1920 channels, resulting in a bandwidth of 468.75 MHz for spw25 and spw27, and 1875 MHz for spw29 and spw31.
The observational data used in our survey covers a total frequency bandwidth of 4.68 GHz.
The spectra have been smoothed to a uniform spectral resolution of 0.976 MHz across all spectral windows.
To improve the signal-to-noise ratio of the spectral images, the solutions obtained from continuum self-calibration were applied to the spectral line data. 

\subsection{The sample of dense cores} \label{sec:core catalogue}
By using the continuum data of the 11 protoclusters, \cite{Xu2024ApJS..270....9X} identified 248 continuum dense cores. The physical parameters (such as core mass, peak flux, H$_{\rm 2}$ column density, number density, surface density, and core radius) were estimated and provided in their paper.
The idenfitited cores are classified into 6 types in \cite{Xu2024ApJS..270....9X}.
In this study, we used the dense core sample reported by them, but we classified all the samples into two categories: prestellar core candidates and protostellar cores based on specific criteria. Cores without any signs of star formation were classified as prestellar core candidates. Cores already associated with star-forming activities or showing signs of warming up were all classified as protostellar cores. The spatial distributions of the dense cores are shown in the left-most panels of Figure~\ref{fig:cores} as blue ellipses, with labeled core names.

\section{Results} \label{sec:results}
\subsection{Detections of S-bearing species}\label{sec:detection}
\begin{figure*}
	\plotone{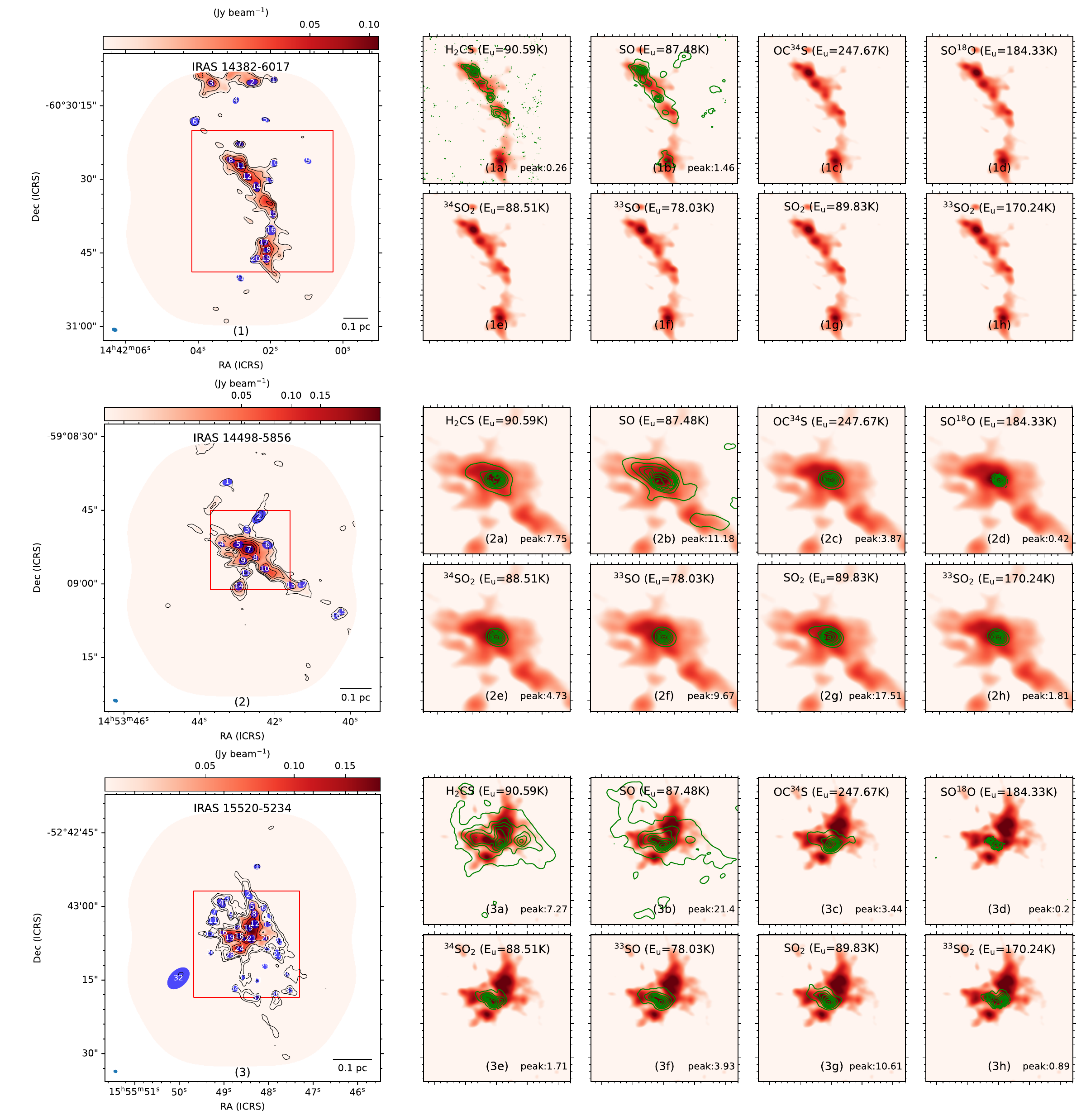}
	\caption{Illustration of dense core locations, continuum emissions, and S-bearing molecular gas distributions. The leftmost sequence of panels, numbered (1) to (11), showcases the continuum emissions, depicted with a red background and outlined in black contours. Contour levels are set at 6$\sigma$, 12$\sigma$, 24$\sigma$, 48$\sigma$, and continue until reaching the peak intensities. The rms noises $\sigma$ are listed in Table \ref{tab:observation}, alongside the locations of dense cores, indicated by numbered blue ellipses. The small panels, such as (1a) to (1h), display the distributions of S-bearing species, represented by green contours. Contour levels range from 5\% to 95\% of peak values in 20\% increments. The dust continuum emissions are superimposed as red backgrounds..	Gas distribution is generated by emissions with intensities 3 times higher than the rms noise. To ensure a clear representation of the gas distributions, the small panels are presented as magnified views. The red rectangles in the leftmost panels correspond to the areas that have been enlarged for detailed examination.}\label{fig:cores}
\end{figure*}

\addtocounter{figure}{-1}
\begin{figure*}
	\plotone{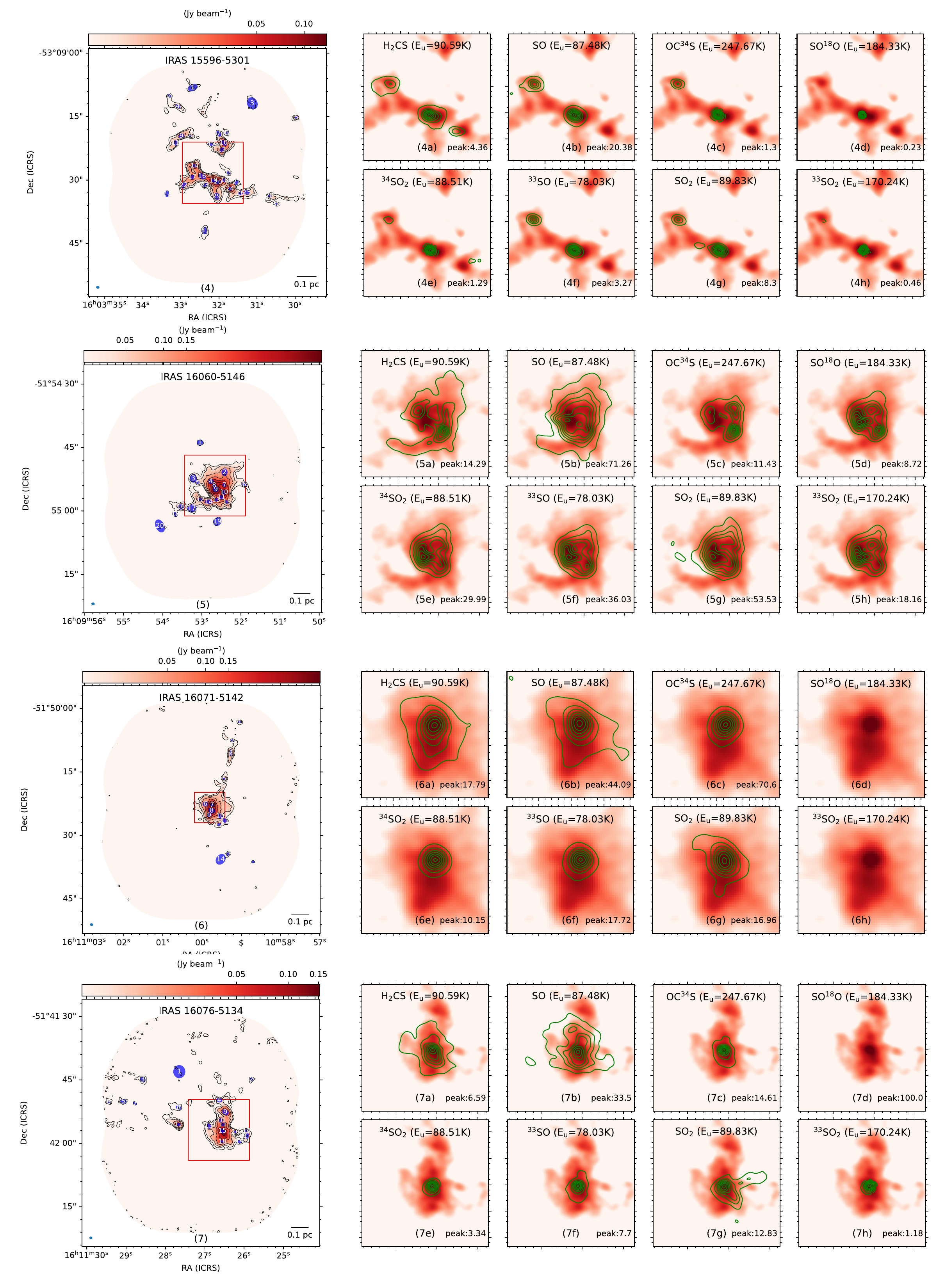}
	\caption{-Continued}\label{}
\end{figure*}

\addtocounter{figure}{-1}
\begin{figure*}
	\plotone{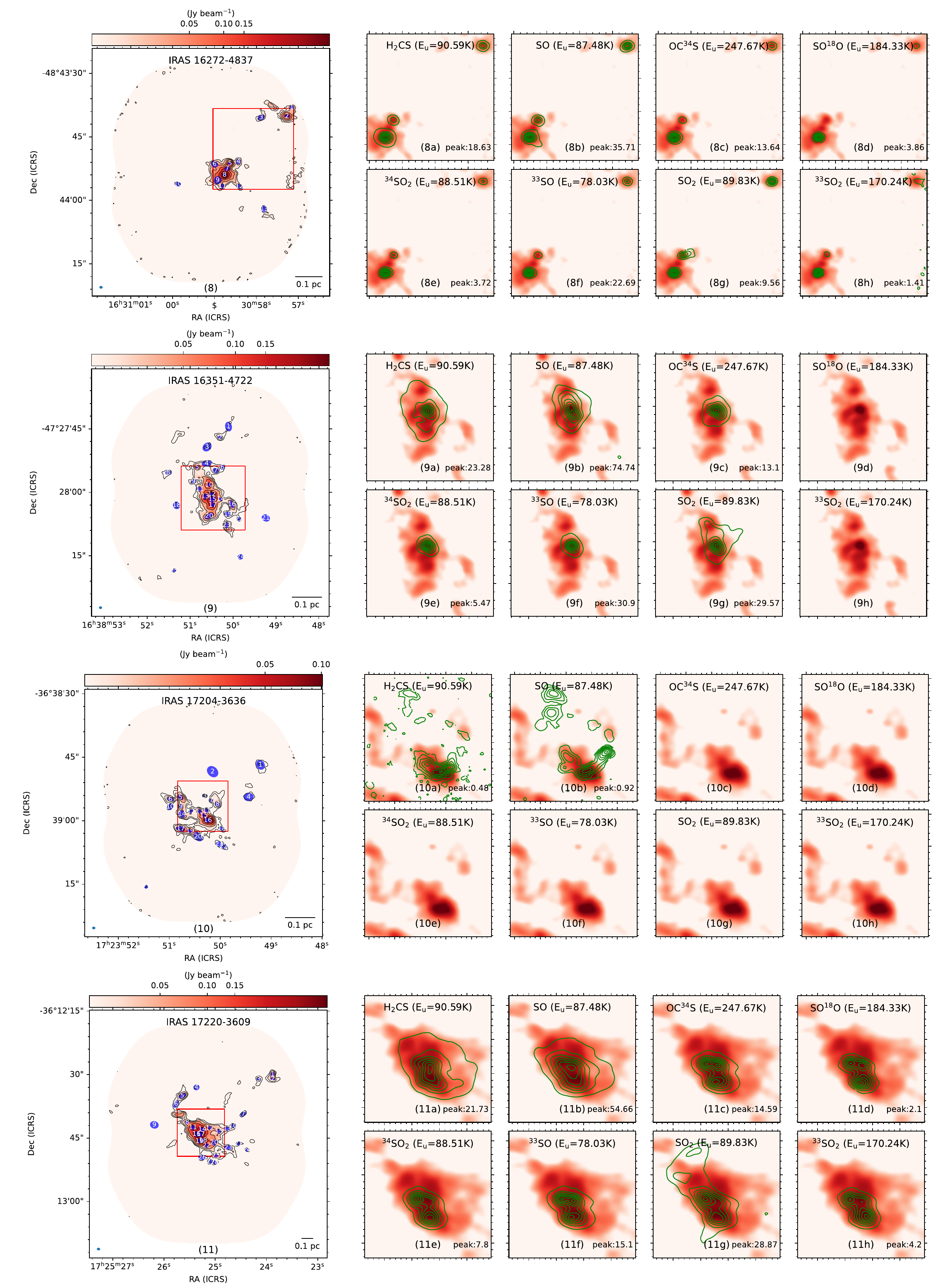}
	\caption{-Continued}\label{}
\end{figure*}

We extracted spectra from the position of the brightest pixel of continuum emissions within the core's region. Line surveys were conducted to identify S-bearing species in dense cores based on the following criteria.
Firstly, the line emission should have at least three continuous spectral channels with intensity exceeding three times the root-mean-square (rms) noise of spectrum.
Secondly, the velocity of the observed transitions must align with the systemic velocity of protocluster.
By checking the spectra of 248 dense cores, a total of 10 S-bearing species (CS, \htcs, SO, \ttso, \sot, \ttsot,  \tfsot,  \seoo,  \octfs,  and NS)  were detected in our observations.
The basic informations of detectable transitions is listed in Table~\ref{tab:transitions}, which are taken from CDMS \cite[]{CDMS}.
It should be noted that not all the transitions listed in the table can be detected in each core due to line blending effect and the different physical environments of the core.

\begin{figure*}
	\plotone{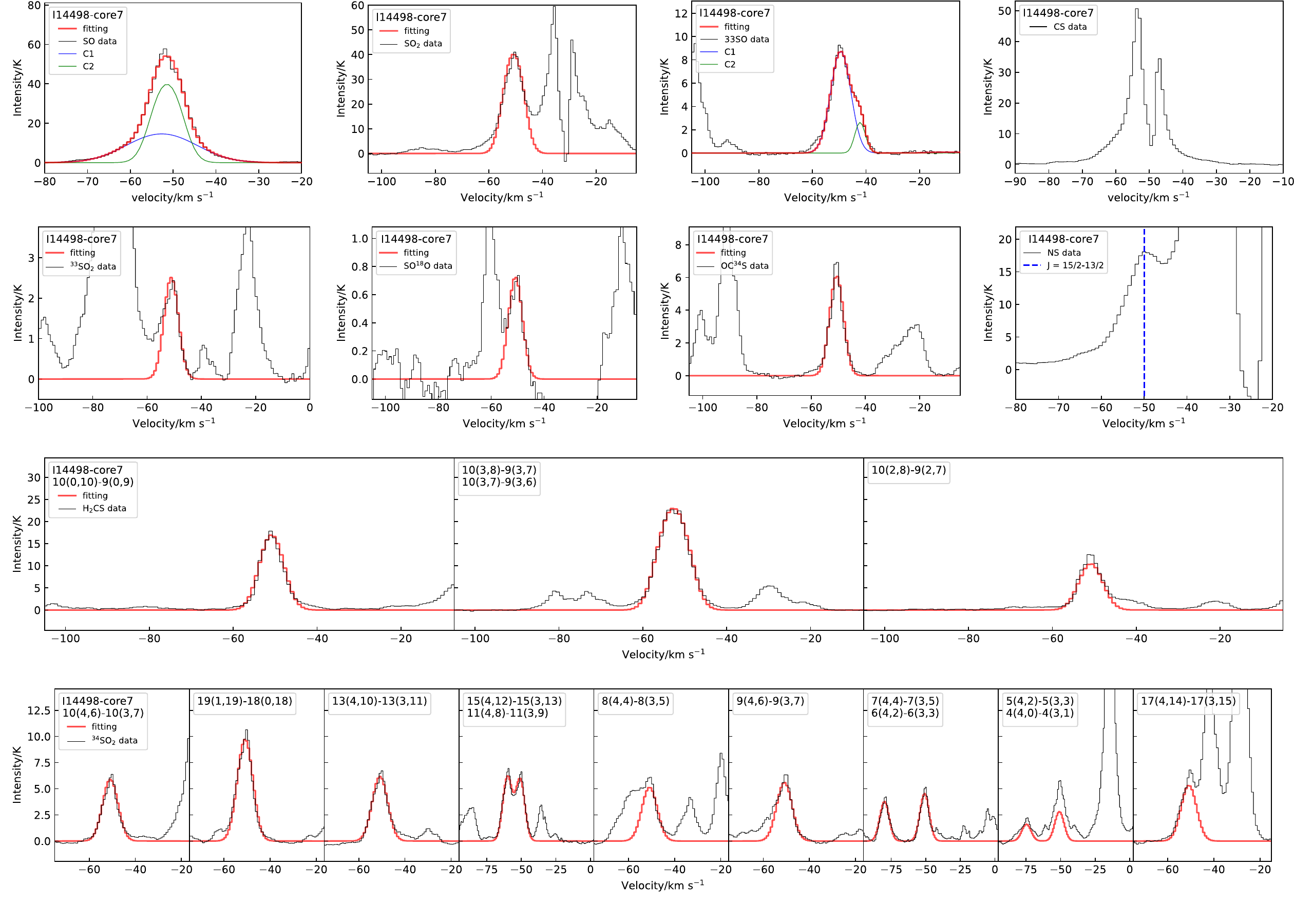}
	\caption{Spectra of the I14498-7 core are depicted. The observation is represented by the black stepped line in each panel. 
		The spectrum exhibiting a non-Gaussian profile has been fitted with multiple velocity components, and the different components are shown as colored curves in each panel. 
		The red stepped line represents the aggregated modeling of all velocity components. 
		Molecular names and component numbers are indicated in the upper left corner of each panel. 
		Multiple transitions of \htcs~and \tfsot~ are also plotted. 
		The quantum numbers (QNs) of transitions for H$_{2}$CS and $^{34}$SO$_{2}$ are presented in the panels as well. The spectra of CS and NS were also presented with only the black stepped line, it is clear that the CS spectrum is double-peaked and with obvious outflow wings, and the NS spectrum is blended by CO outflow wings. Therefore, the spectra of CS and NS are excluded from the analyses.} 
	\label{fig:spectra}
\end{figure*}

Figure~\ref{fig:spectra} presents the spectra of the I14498-7 core as an illustrative example. For \ttsot, \seoo, and \sot, multiple detectable transitions fall within the observed frequency range, only one transition of each species that is less affected by line blending effects is shown in Figure \ref{fig:spectra}. In the case of \tfsot, all detectable transitions are covered by spw29, and most of them are less affected by the line blending effect. Therefore, we present all transitions of \tfsot~in Figure \ref{fig:spectra}. A total of 8 detectable transitions of \htcs~exist, with only 4 unblended transitions shown in the figure.
	
The spectra of CS generally present obvious optically thick absorption dips and outflow wings, and the spectra of NS are significantly blended with CO outflow wings. We detected the CS and NS emissions, but the useful parameters are hardly derived from their spectra. Therefore, we presented the observed spectra of CS and NS without fitting results, and excluded them from further analyses. The S-bearing molecular spectra of all dense cores can be found in Appendix~\ref{sec:B} as figure sets.

\begin{figure}
	\plotone{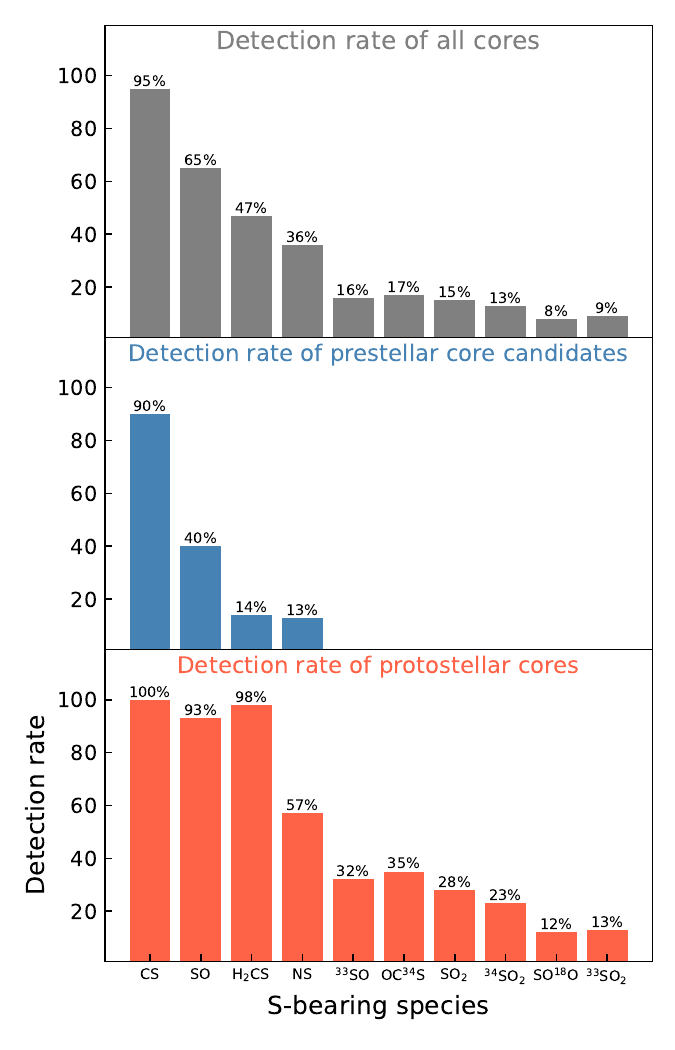}
	\caption{The detection rate of S-bearing species. The upper, middle, and bottom panels present detection rates of all cores, prestellar core candidates, and protostellar cores, respectively.} \label{fig:detection rate}
\end{figure}

Figure~\ref{fig:detection rate} illustrates the detection rates of different molecules. 
The number of detected species for each core is listed in the last column of Table~\ref{tab:catalog}. 
The majority of dense cores (236 cores in total) detected at least one S-bearing species, these cores are labeled with number ``1'' to``10'' in the last column of Table \ref{tab:catalog}.
A total of 12 dense cores do not have any detection of S-bearing species, as abeled with ``0'' in the last column of Table \ref{tab:catalog}. 
There are 10 dense cores detected 10 S-bearing species, as labeled with ``10'' in the last column of Table \ref{tab:catalog}.
\cite{Artur2023A&A...678A.124A} reported the detection rates of 90\% of CS, 78\% of SO, and 44\% of \sot~in a sample of 50 Class 0/I protostellar sources in the Perseus molecular cloud. 
Our observations show a comparable detection rate for CS and SO in protostellar cores with theirs. The \sot~detection rate is slightly lower because the \sot~emissions in some cores are blended with HCO$^{+}$(4-3), preventing us from claiming the detection of \sot~in these cores.

The integrated intensity maps (moment 0) for the detected species were generated and presented as green contours within the small panels of Figure \ref{fig:cores}.
In the process of generating these maps, we have excluded emission that falls below three times the root mean square (rms) noise level. From Figure \ref{fig:cores}, gas distributions of \htcs~and SO exhibit a notable similarity and are more spatially extended compared to those of other species. The distributions of the species, including \sot, \octfs, \ttso, \tfsot, \ttsot, and \seoo, are found to be analogous to one another with compact morphologies.
The distribution of \sot~are similar to the isotopologues but with a more pronounced extension, which attributed to the partly blending with the HCO$^{+}$ line.  The extended \sot~features in the gas distribution are identified as contamination from the outflow wings of HCO$^{+}$.

\subsection{Estimation of molecular parameters}\label{sec:parameter}
\subsubsection{Parameters derived from emissions}\label{sec:emission}
We employed the \texttt{SLIM} tool from the \texttt{MADCUBA} package\footnote{\url{https://cab.inta-csic.es/madcuba/}} \cite[][]{2019A&A...631A.159M} to interactively estimate the physical parameters and their associated uncertainties under local thermodynamic equilibrium (LTE) assumption. 
The \texttt{MADCUBA} has garnered widespread application in the field of star formation \cite[e.g.][]{2023A&A...677A..15M,2021ApJ...920L..27Z,2021ApJ...917...44J,2020A&A...641A..54C,2020MNRAS.496.1990R}.
The parameters were derived by fitting an LTE model on the observed spectral emissions. The detailed fitting processes are presented in Appendix \ref{sec:A}. 
The LTE assumption is pivotal for parameter estimation. To ascertain the validity of the LTE assumption for our observations, we calculated the critical densities for SO, \htcs, and \sot~using collisional data from the LAMDA database \cite[][]{van2020Atoms...8...15V} at a uniform kinetic temperature of 30 K\footnote{The average dust temperature of 11 protoclusters is 28 K. Therefore, we adopted 30 K to estimate critical densities}. The critical densities were determined to be 3.7$\times$10$^{3}$, 8.3$\times$10$^{4}$, and 7.5$\times$10$^{5}$ cm$^{-3}$, respectively. From Table \ref{tab:catalog}, only one dense core, I15520-core32, exhibits a molecular hydrogen number density ($n_{H_2}$) below 10$^{5}$ cm$^{-3}$. Consequently, for SO, \htcs, and \sot, the LTE assumption is deemed reliable. 
We were unable to estimate the critical densities for \ttso, \ttsot, \tfsot, \octfs, and \seoo~due to the lack of collisional data. \cite{Goldsmith2001ApJ...557..736G} suggested that dust and gas are well coupled at molecular hydrogen densities exceeding 10$^{6}$ cm$^{-3}$. In our sample, almost all the dense cores\footnote{Only I15520-core4, I115520-core21, I115520-core34, I16060-core2, I16060-core10, I17220-core20, and I17220-core22 have $n_{\rm H_{2}}$ lower than 10$^{6}$ cm$^{-3}$.} with S-bearing species detections have $n_{H_2}$ higher than 10$^{6}$ cm$^{-3}$. Given these conditions, the LTE assumption in our study is considered reliable for all species in the majority of dense cores.

\htcs~and \tfsot~exhibit multiple unblended transitions that span a wide range of upper energy levels (see Table~\ref{tab:transitions}).
We priorly conducted LTE fitting on the spectra of \htcs~and \tfsot~to ascertain the gas temperatures. These specific transitions have also been utilized for temperature estimation in other studies \cite[e.g.][]{Liu2011ApJ...730..102L,Chen2024ApJ...962...13C}.
Out of the 115 cores that exhibited detectable \htcs~transitions, 99 of them possessed a minimum of four transitions suitable for the derivation of gas temperatures. For the \tfsot~transitions, detections were made in 27 cores, and all were found to be adequate for calculating excitation temperatures.

It should be noted that 4 transitions of \htcs~less affected by line blending effect were used to estimate the rotational temperature. 
We marked these transitions with ``$\ast$'' in Table~\ref{tab:transitions}. 
\tfsot~exhibits 12 detectable transitions in the observed data, and these transitions are not significantly affected by the line blending effect. 
This characteristic renders \tfsot~a reliable tracer of gas temperature for S-bearing species. 
However, the low detection rate of 13\% suggests that it can only be utilized to trace the temperatures of a limited number of dense cores.

Due to the absence of multiple unblended transitions for the species SO, \ttsot, \octfs, \sot, \seoo, and \ttsot, their respective temperatures could not be determined through fitting processes. As elaborated in Section \ref{sec:parameter}, the gas distributions of \htcs~and SO are analogous, and a similar pattern is observed in their abundances, as detailed in Section \ref{sec:H2CS&SO}. Therefore, we assume that \htcs~and SO are likely to emanate from analogous environments, and may have similar gas temperatures. With this assumption, we utilized the gas temperature derived from \htcs~as a fixed parameter in the LTE fitting for the SO spectra. Consequently, only the column density, systemic velocity, and line width were treated as free parameters in the fitting process for SO.
	
Similarly, the gas distributions of \sot, \ttso, \octfs, \ttsot, and \seoo~are analogous to those of \tfsot. Based on this similarity, we hypothesize that these species also originate from comparable environments and exhibit similar temperatures. Accordingly, the temperature obtained from \tfsot~was adopted as a fixed parameter in the LTE fitting for their respective spectra, with only the column density, systemic velocity, and line width serving as free parameters for the fitting.
	
It is important to note that the assumption of similar origins for the S-bearing species introduces additional uncertainties into the derived physical parameters. However, in the absence of superior alternatives, this approach is currently the most viable for estimating the parameters of SO, \ttso, \octfs, \sot, \seoo, and \ttsot. The dust temperatures of 11 massive protoclusters were given by \cite{Faundez2004A&A...426...97F} and \cite{Urquhart2018MNRAS.473.1059U}, but their angular resolutions are about 20 to 30 times larger than ours. They cannot be directly applied to estimate the parameters of the species under investigation without causing significant uncertainty.

During the fitting processes, a Gaussian profile function is generally assumed. It is evident that the spectra of some detected species, SO in Figure \ref{fig:spectra} for instance, exhibit non-Gaussian profiles.
This phenomenon could be attributed to three main reasons.
Firstly, the presence of inhomogeneous/bulk motion of the gas envelope can lead to asymmetric non-Gaussian profiles.
Secondly, fragmentation may occur on scales below our minimum spatial resolution, resulting in substructures with different systemic velocities.
Thirdly, the interferometer is filtering-out extended structures larger than the LAS of the observations.
Therefore, the observed spectra represent a stacking of Gaussian profiles from several velocity components. To properly obtain parameters from spectra, we fitted spectra with multiple velocity components in case that they exhibited obvious non-Gaussian profiles.
Initially, a fitting with a single Gaussian component is carried out on a spectrum. If the fitting residual shows at least three consecutive spectral channels with intensities exceeding three times of rms noise, an additional Gaussian component is incorporated into the model. This iterative process continues until the fitting residual is subdued to below three times the rms noise threshold.
Results of each individual component's fit are graphically represented by distinct colored curves in Figure~\ref{fig:spectra}. 
The aggregated fitting results of all components are shown with a red stepped line, giving a comprehensive overview of the spectral fitting.
    
The parameters extracted from the \texttt{SLIM} fittings are compiled in the tables in Appendix \ref{sec:A}. It is noteworthy that the \texttt{SLIM} fitting accounts for optical depth in its analysis. During the fitting process, only the \htcs~10(0,10)-9(0,9) transition and the SO emissions in some massive cores exhibited optical depths exceeding the value of 1. Most line emissions are optically thin, which implies that the influence of optical depth on our results and subsequent analyses is negligible.

\subsubsection{Column density estimations for SO absorptions}\label{sec:absorption}
The SO spectra of certain cores display discernible absorption features, as illustrated in Figure Set \ref{fig:C1_SO}. Under the LTE assumption, the conventional approach to calculate the column density from SO absorption lines involves an integration over velocity with optical depth. This process is followed by the application of the subsequent equation \cite[][]{Mangum2015}:
\begin{equation}\label{eq:1}
	N=\frac{3h}{8\pi^3 S\mu^2}\frac{Q_{rot}}{g_u}\frac{exp(\frac{E_{u}}{kT_{ex}})}{ exp(\frac{h\nu}{k T_{ex}})-1  }\times \int\tau_{\nu}dv,
\end{equation}
where $k$ and $h$ are Boltzmann and Planck constants, respectively. 
The permanent dipole moment $\mu$, line strength $S$ are taken from CDMS \cite[][]{CDMS},
$g_u$ is level degeneracy.
The partition function $Q_{rot}$ of SO can be estimated by an approximation for linear polyatomic molecules \cite[][]{1988JChPh..88..356M} as:
\begin{equation}
	Q_{rot} \simeq \frac{k T}{h B_0}exp(\frac{h B_0}{3 k T}),
\end{equation}
where $B_{0}$ is rotational constant.

Optical depth for SO absorption can be estimated by the following approximation:
\begin{equation}
	\tau_\nu = -ln \left[  1- \frac{T_R}{f (J(T_{cont})-J(T_{ex}))} \right],
\end{equation}
where $T_{R}$ is radiation temperature of SO absorption, $J(T_{cont})$ is Rayleigh–Jeans equivalent temperature of background continuum emissions, and $J(T_{ex})$ is Rayleigh–Jeans equivalent temperature of $T_{ex}$.
The gas temperatures derived from \htcs~were utilized to calculate the optical depth. In instances where the excitation temperature is not available, the dust temperature listed in Table \ref{tab:observation} will be employed as a surrogate for these calculations. The column densities of SO absorptions are listed in Table \ref{tab:so}.

\subsection{Abundances}\label{abundance}
Column densities of S-bearing species are estimated from the spectra extracted from pixels coinciding with the peak positions of continuum emissions within these dense cores. Consequently, the peak \hmole~column density ($N_{H_2}$) from \cite{Xu2024ApJS..270....9X} is adopted for calculating the relative abundances of S-bearing species to \hmole. 
The peak $N_{\rm H_2}$ and the column densities of S-bearing species were calculated from the same emitting areas.
The resulting abundances of the S-bearing species for each core are detailed in Table \ref{tab:catalog}.
It is important to note that the abundances listed in Table \ref{tab:catalog} are computed from the integrated column density across all velocity components. As such, each species within a dense core is associated with a singular abundance value, irrespective of the number of velocity components detected.

The estimation of $N_{\rm H_2}$ may be influenced by the optical depth of the continuum emission, given that the initial estimate was based on the assumption of an optically thin continuum emission \cite[][]{Xu2024ApJS..270....9X}. To ascertain whether the continuum emissions are optically thin or thick, we have calculated the optical depth of the continuum emission employing the following equation \cite[][]{Frau2010ApJ...723.1665F,Gieser2021A&A...648A..66G}: 

\begin{equation}
	\tau_{\rm cont} = ln[1-\frac{S_{\nu}}{\Omega B(T_{\rm d})}]
\end{equation}

In the equation, $S_{\nu}$ represents the flux density at the continuum peak position of the core, $\Omega$ denotes the solid angle of the beam, and $B(T_{\rm d})$ is the Planck function evaluated at the dust temperature $T_{\rm d}$. Utilizing the dust temperatures listed in Table~\ref{tab:observation}, we estimated the optical depth from $10^{-5}$ to $10^{-2}$. These values suggest that the estimations of $N_{\rm H_2}$ are in line with the optically thin assumption for the continuum emission. 
\startlongtable
\begin{deluxetable*}{cccccccccccccc} 
	\setlength{\tabcolsep}{0.02in}
	\label{tab:catalog}
	\tabletypesize{\tiny}
	\tablecaption{Physical properties and molecular abundances of the dense cores}
	\tablewidth{0pt} \tablehead{
		\colhead{Core$^{(a)}$}  &\colhead{Position}    &\colhead{Mass$^{b}$}  &\colhead{$N_{\rm H_{2}}$$^{b}$}   &\colhead{$n_{\rm H_{2}}$$^{b}$}   &$f_{\rm SO}^{\ast}$   &$f_{\rm H_{2}CS}$ &$f_{\rm ^{33}SO}^{\ast}$ &$f_{\rm OC^{34}S}^{\ast}$ &$f_{\rm SO_2}^{\ast}$ &$f_{\rm ^{34}SO_{2}}$ &$f_{\rm SO^{18}O}^{\ast}$ &$f_{\rm ^{33}SO_{2}}^{\ast}$   &Num\\
		\colhead{}                 &\colhead{(R.A., Decl)}        & &\colhead{10$^{22}$}     &\colhead{10$^{6}$}     &\colhead{(10$^{-10}$) } &\colhead{(10$^{-10}$) }   &(10$^{-10}$)  &(10$^{-10}$)  &(10$^{-10}$)   &(10$^{-10}$)  &(10$^{-10}$)  &(10$^{-10}$)      &      \\
		&(degree) &\colhead{(M$_\odot$)}  &(cm$^{-2}$) &(cm$^{-3}$)
	}
\startdata
I14382-1$^{(0)}$  &220.5079675,-60.5026951 &1.3(0.9)	&3.7(2.4)	 &9.4(3.4)	    &--          &--          &--           &--           &--           &--          &--        &--        &1  \\
I14382-2$^{(0)}$  &220.5104280,-60.5028481 &3.2(2.1)	&0.9(0.6)	 &14.6(6.0)	    &0.4(0.2)    &--          &--           &--           &--           &--          &--        &--        &2  \\
I14382-3$^{(0)}$  &220.5151380,-60.5028925 &2.3(1.6)	&6.4(4.4)	 &19.2(7.8)	    &--          &--          &--           &--           &--           &--          &--        &--        &1  \\
I14382-4$^{(0)}$  &220.5123067,-60.5038647 &0.6(0.4)	&1.6(1.2)	 &4.6(2.5)	    &2.7(1.5)    &--          &--           &--           &--           &--          &--        &--        &2  \\
I14382-5$^{(0)}$  &220.5089381,-60.5049506 &0.5(0.3)	&1.4(0.9)	 &3.9(1.6)	    &--          &--          &--           &--           &--           &--          &--        &--        &1  \\
I14382-6$^{(0)}$  &220.5170375,-60.5050743 &0.6(0.4)	&0.1(0.1)	 &2.8(1.5)	    &--          &--          &--           &--           &--           &--          &--        &--        &1  \\
I14382-7$^{(0)}$  &220.5118667,-60.5063203 &1.7(1.1)	&4.9(3.2)	 &13.4(5.0)	    &--          &--          &--           &--           &--           &--          &--        &--        &1  \\
I14382-8$^{(1)}$  &220.5129036,-60.5072444 &2.0(1.4)	&5.6(3.8)	 &13.0(3.8)	    &0.6(0.2)    &0.6(0.2)    &--           &--           &--           &--          &--        &--        &3  \\
\enddata
	\tablecomments{Columns 6 to 13 represent the fractional abundance related to H$_{\rm 2}$. The last column is the number of detected species, the detections of CS and NS were included because although their column densities cannot be derived the detection of their transitions is reliable.\\
	$\ast$: The abundances of the species were estimated by fitting only one unblending transition, for these species, the gas temperatures are set as fixed parameters during fitting.\\
	In some cores (16060-7, 16060-16, 16060-18, 16071-7, 16272-5), isotopologues of \tfsot, \ttsot, and \seoo~were detected and their abundances were listed. The abundances of main isotopologue (\sot) are absent, due to the fact that \sot~emissions are blended with HCO$^{+}$(4-3) outflow line wings, preventing abundance estimations for \sot~in these cores.}
	\tablenotetext{a}{: The numbers indicate the classification of cores, 0: prestellar core candidates, 1: protostellar cores.}
	\tablenotetext{b}{: The parameters of mass, $N_{\rm H_2}$, and $n_{\rm H_2}$ are taken from \cite{Xu2024ApJS..270....9X}.}
\end{deluxetable*}

\section{Discussion} \label{sec:discussion}
\begin{figure*}
	\plotone{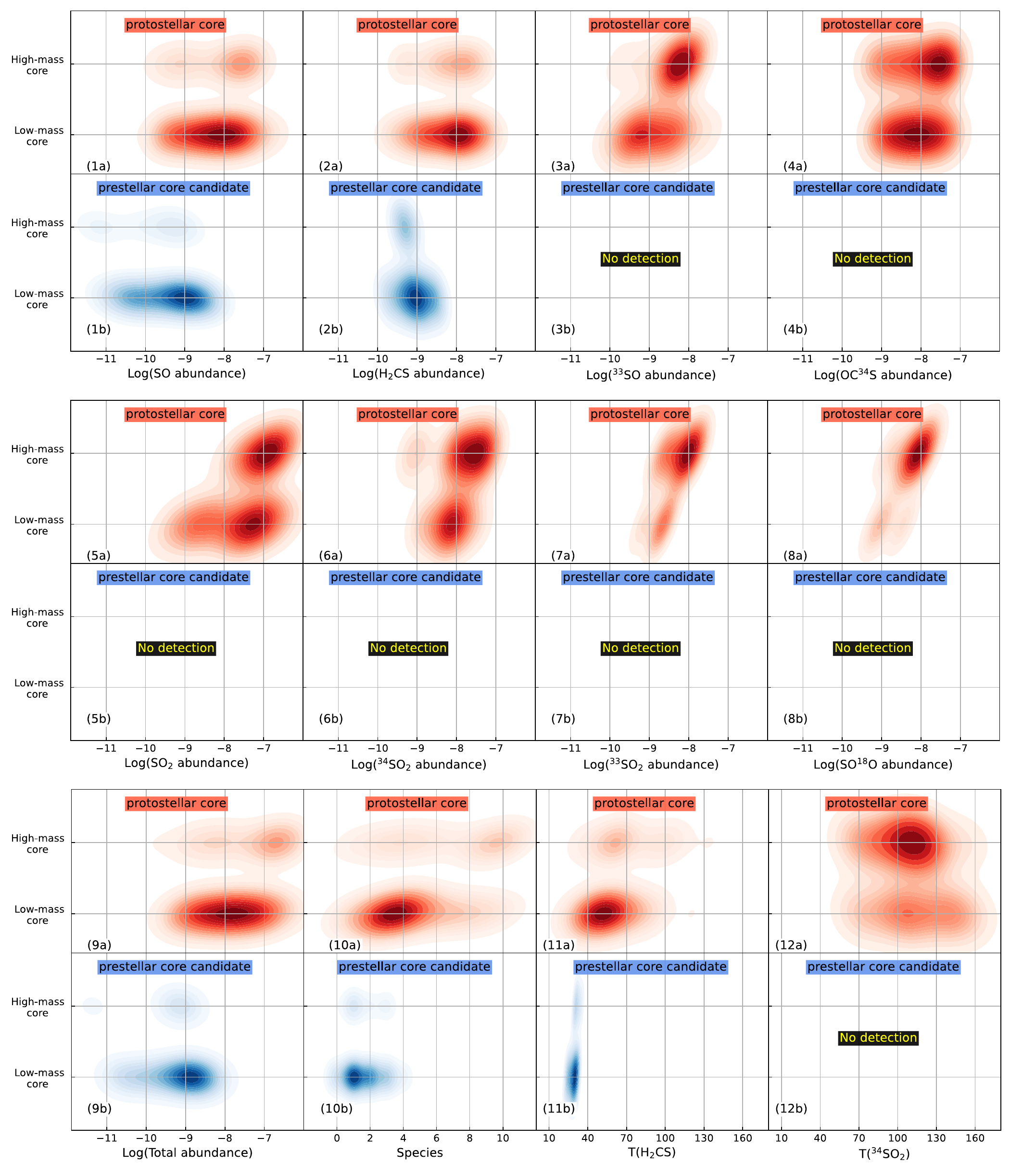}
	\caption{Two-dimensional KDE plots of molecular abundances, number of detected species, and temperatures versus low- and high-mass core categories. Prestellar core candidates and protostellar cores are analyzed separately and shown in blue and red colors, respectively. The vertical axis in each panel shows the mass categories of low- and high-mass cores, while the horizontal axis in each panel displays a specific parameter used for KDE analysis. Some species such as \ttso, \octfs, \sot, \tfsot, \ttsot, and \seoo~were not detected in prestellar core candidates, labeled as ``No detection" in the panels.} 
	\label{fig:kde_2D}
\end{figure*}

The sample of 248 dense cores covers a wide range of densities ($N_{H_2}$:2.7$\times$10$^{22}$ $\sim$ 2.5$\times$10$^{24}$ cm$^{-2}$,
$n_{H_2}$:1.0$\times$10$^{4}$ $\sim$ 3.7$\times$10$^{8}$ cm$^{-3}$), mass (0.1 $\sim$ 52.6 M$_{\odot}$), temperature ($T_{\rm H_2CS}$: 39.1 $\sim$ 134.7 K; $T_{\rm ^{34}SO_{2}}$: 72.3 $\sim$ 158 K), molecular abundance (total abundance: 4.4$\times$10$^{-12}$ $\sim$ 6.9$\times$10$^{-7}$), and evolutionary stages (prestellar core candidates and protostellar cores).
Statistical analyses of these data provide an opportunity to investigate the properties of S-bearing species in dense cores, which cannot be obtained through individual case studies alone. 
In this section, we will primarily conduct a statistical analysis of the parameters we have obtained to glean more information regarding the characteristics of S-bearing species, as well as discuss the similarities and differences in gas distributions, abundance, and gas temperatures among various S-bearing molecules, culminating in a discussion on the issue of sulfur depletion.

\subsection{Similarities and differences between \htcs~and SO}\label{sec:H2CS&SO}
Kernel density estimation (KDE) is a non-parametric technique that employs kernel smoothing to infer the probability density function of a dataset. This method stands out as a valuable tool for visualizing the distribution of data. Unlike histograms, which may be subject to the arbitrariness of bin width selection, KDE provides a more fluid representation of the underlying data patterns.
We categorized core samples into distinct groups: low-mass, high-mass prestellar core candidates and protostellar cores. For each category, we performed separate KDE analyses to elucidate the distribution characteristics. Figure~\ref{fig:kde_2D} presents the two-dimensional KDE plots, which illustrates parameters of molecular abundances, total abundance, number of detected species, and excitation temperatures for \htcs~and \tfsot, respectively.
	
Among the S-bearing species, a notable distinction is found: SO and \htcs~are present in both prestellar core candidates and protostellar cores, whereas other species are exclusively detected in protostellar environments.
As depicted in panels (1a) to (2b) of Figure \ref{fig:kde_2D}, the abundances of SO and \htcs~exhibit a common trend, with prestellar core candidates displaying lower abundances on the order of approximately 10$^{-9}$, and protostellar cores exhibiting higher abundances around 10$^{-8}$.  This suggests a potential enhancement in both SO and \htcs~abundances as dense cores evolve. The high detection rates of SO and \htcs~indicate that these two species span various evolutionary stages from prestellar to protostellar cores \cite[][]{Chen2024ApJ...962...13C}. 
The gas distributions of SO and \htcs~in Figure \ref{fig:cores} demonstrate similar morphologies, underscoring a significant resemblance between SO and \htcs.

While SO and \htcs~exhibit several similarities, they also display notable differences. One key distinction lies in their detection rates, as depicted in Figure \ref{fig:detection rate}. The higher detection rate of SO may suggest that it forms more easily in prestellar cores compared to \htcs. This enhanced formation could be due to the efficient production of SO through barrierless neutral-neutral reactions in the gas phase \cite[][]{Fuente2016A&A...593A..94F,Semenov2018A&A...617A..28S,Laas2019A&A...624A.108L,Taquet2020A&A...637A..63T,Booth2023A&A...669A..53B}. Furthermore, SO has been recognized as a tracer of diverse physical conditions, including protoplanetary disks, planet formation regions, shock zones, outflows, hot cores \cite[e.g.][]{Sakai2014Natur.507...78S,Sakai2017MNRAS.467L..76S,Tabone2017A&A...607L...6T,Lee2018ApJ...863...94L,Taquet2020A&A...637A..63T,vanGelder2021A&A...653A.159V,Tychoniec2021A&A...655A..65T}. 

The KDE plot of SO abundances reveals an extension to lower values, reaching approximately 10$^{-11}$, while \htcs~abundances are at higher levels, above 10$^{-10}$. Our results support the notion that SO may be formed in more diffused regions compared to \htcs. The abundance range of \htcs~reported in other similar studies, from 1.2$\times$10$^{-10}$ to 5$\times$10$^{-8}$ \cite[e.g.][]{Minh2011ApJ...737L..25M,Shimajiri2015ApJS..221...31S,Minh2018ApJ...864..102M,Luo2019ApJ...885...82L,Rivi2019A&A...628A..16R,Fuente2021MNRAS.507.1886F,Moller2021A&A...651A...9M,Bouscasse2022A&A...662A..32B,Chen2024ApJ...962...13C,Fontani2023A&A...680A..58F}, is consistent with the our results. There are also regions detected \htcs~with abundance as low as $\sim$10$^{-11}$ in PDRs \cite[][]{Rivi2019A&A...628A..16R}, where is quiet different with our sources.

In summary, SO demonstrates a propensity for detection across a wider range of physical and chemical environments although numerous similarities in the properties of SO and \htcs~are found.

\subsection{\htcs~and \tfsot~trace environments with different temperatures}\label{sec:H2CS&34SO2}
Panel (11a) to (12b) of Figure \ref{fig:kde_2D} present the two-dimensional KDE plots of \tone~and \ttwo~in prestellar core candidates and protostellar cores.  
The analysis reveals that the kernel density peaks for \tone~fall below 70 K across both low- and high-mass cores. In contrast, the peaks for \ttwo~are distinctly situated at temperatures exceeding 100 K for cores of both mass categories.
	
The KDE plot of \htcs~shows an extension towards temperatures as high as 130 K in high-mass cores, which can be attributed to the presence of some hot cores in our samples \cite[][]{Liu2023ApJ...958..174L,Chen2024ApJ...962...13C,Xu2024ApJS..270....9X}, and they generally have gas temperatures exceeding 100 K.
Comparing to the KDE plot of \tone, the kernel density for \ttwo~extends to even more elevated temperatures (\textgreater150 K) for both low- and high-mass cores.
This indicates that \tfsot~and \htcs~trace the regions with different temperatures in both low- and high-mass cores.
The Spearman correlation coefficient between \tone~and \ttwo~was calculated as 0.57, indicating a merely moderate positive correlation between them, also suggesting that \htcs~and \tfsot~seem to trace gases with different temperatures. 
	
\cite{Fuente2021MNRAS.507.1886F} investigated the hot core in Mon R2, noted a spatial offset between the distribution of \htcs~gas and the continuum peak, suggesting that \htcs~may not be an accurate tracer of the hot core.
The gas distributions of \htcs~and \tfsot~illustrated in Figure \ref{fig:cores} demonstrate that the \htcs~is more extended than \tfsot. Additionally, the presence of \htcs~is not invariably associated with hot cores/corinos\footnote{The identification of hot core/corino can be found in \cite{Xu2024ApJS..270....9X}}, but the majority of \tfsot~detections ($\sim$89\%) were indeed found in association with hot cores/corinos.
Consequently, our results indicate that \htcs~appears to trace the outer warm envelope of the core \cite[][]{Fuente2021MNRAS.507.1886F,Tychoniec2021A&A...655A..65T,Bouscasse2022A&A...662A..32B}, while \tfsot~is more representative of the high-temperature central regions.

\subsection{The differences between high-mass and low-mass cores}\label{sec:low&high}
Panels (3a) to (8b) in Figure \ref{fig:kde_2D} present that \ttso, \octfs, \sot, \ttso, \octfs, \tfsot, \ttsot, and \seoo~were exclusively found in protostellar cores, with their abundances notably differing between high-mass and low-mass cores. Specifically, \ttso, \tfsot, \ttsot, and \seoo~display a higher prevalence and abundances in high-mass cores, implying an enhancement of these species in high-mass star-forming regions. 
Additionally, within protostellar cores, both the diversity of detected species and the total molecular abundance are more pronounced in high-mass cores. Most strikingly, the dense cores that exhibit detections of all S-bearing species are predominantly high-mass, accounting for 80\%.
Our findings suggest that the process of high-mass star formation is associated with an enhancement in both the diversity of S-bearing species and the abundance.
	
In the case of \sot, the kernel density peaks for both low-mass and high-mass cores are approximately 10$^{-7}$. \sot~has been observed in low-mass cores with abundances as low as 10$^{-10}$, while in high-mass cores, the majority exhibit \sot~abundances exceeding 10$^{-8}$, with a kernel density peak centered around 10$^{-7}$. 
	
\begin{figure}
	\plotone{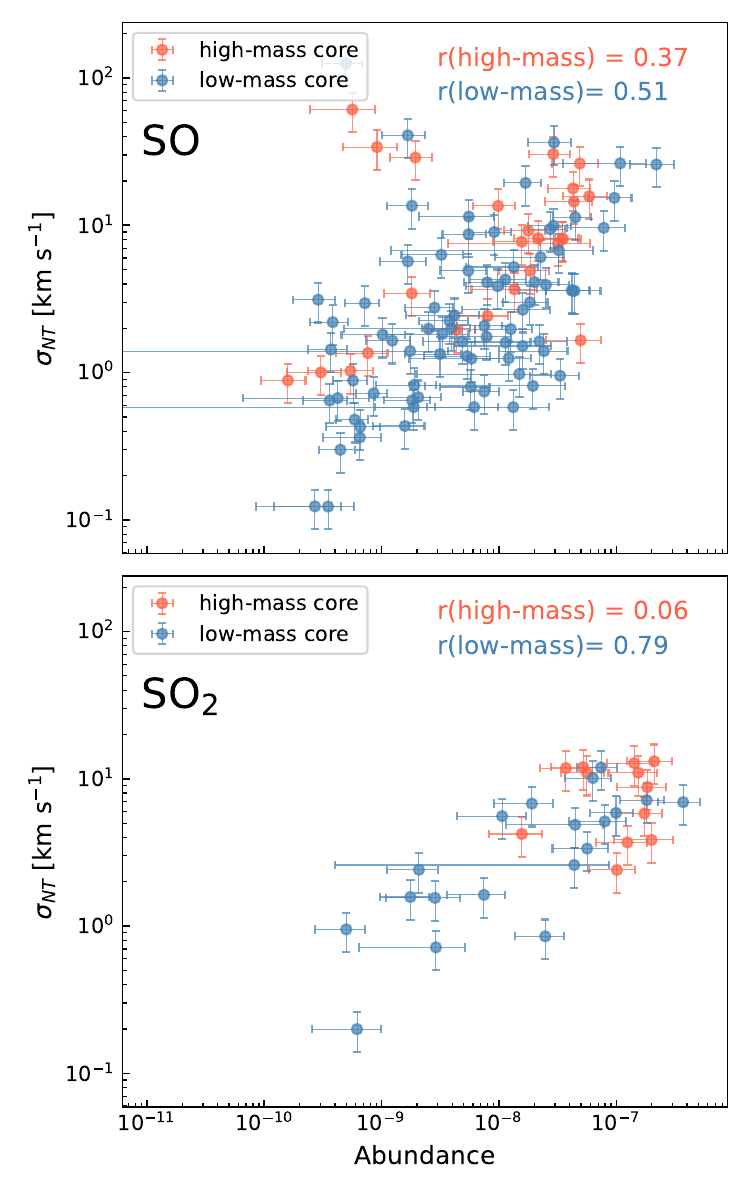}
	\caption{The correlation between the non-thermal velocity dispersions ($\sigma_{NT}$) of SO (upper panel) and \sot~(bottom panel) versus their abundances. In each panel, high-mass cores and low-mass cores are shown as orange and blue colors, respectively. The Spearman correlation coefficient ($r$) is displayed in the upper-right corner of each panel.} 
	\label{fig:sigma_ab}
\end{figure}

It is widely recognized that SO and \sot~are potential tracers of shocked gas, with their abundances potentially enhanced by the effects of shocks \cite[][]{Sakai2014Natur.507...78S,Sakai2017MNRAS.467L..76S,Artur2019A&A...626A..71A,Oya2019ApJ...881..112O,vanGelder2021A&A...653A.159V,Tychoniec2021A&A...655A..65T}.
To ascertain whether the abundances of the observed S-bearing species were influenced by shocks or other non-thermal processes, we calculated the non-thermal velocity dispersion ($\sigma_{NT}$) for identified species as following equation:

\begin{equation}
	\sigma_{NT} = [\sigma_{obs}^2-\frac{k T_{ex}}{m}]^{\frac{1}{2}},
\end{equation}

where $\sigma_{obs} = v_{FWHM}/\sqrt{8ln2}$ is the observed total velocity dispersion, and $v_{FWHM}$ is line width from tables in Appendix \ref{sec:A}. $k$ is Boltzmann constant, and $m$ is the molecular mass.
The relationship between $\sigma_{NT}$ and their abundances are investigated and shown in Figure \ref{fig:sigma_ab}. The figure presents each panel with the Spearman correlation coefficients, where high-mass cores and low-mass cores are denoted by orange and blue, respectively. 
	
For SO, the Spearman correlation ($r$) values of 0.37 and 0.51 in high-mass and low-mass cores suggest that SO abundance is not significantly enhanced by non-thermal motions. Concerning \sot, with $r$ values of 0.06 and 0.79 in high-mass and low-mass cores, indicating non-thermal motions appear to well correlated with \sot~abundance in low-mass cores. 
The robust correlation of low-mass cores suggests that the abundances of \sot~can be enhanced by non-thermal motions, and a significant subset of low-mass cores remains in relatively quiescent environments. 
In high-mass cores, no discernible correlation exists between \sot~abundances and $\sigma_{NT}$. However, these cores are clustered in the upper-right corner of the panel, indicating a general trend of higher $\sigma_{NT}$ levels coincident with increased \sot~abundance across all high-mass cores.
	
Our results suggest that high-mass cores likely exhibit more vigorous star-forming activities and potentially higher-velocity shocks, resulting in enhanced \sot~abundances across all high-mass cores. 
	
\subsection{S-bearing molecular abundances enhanced by outflows}\label{sec:outflow}
\begin{figure}
	\plotone{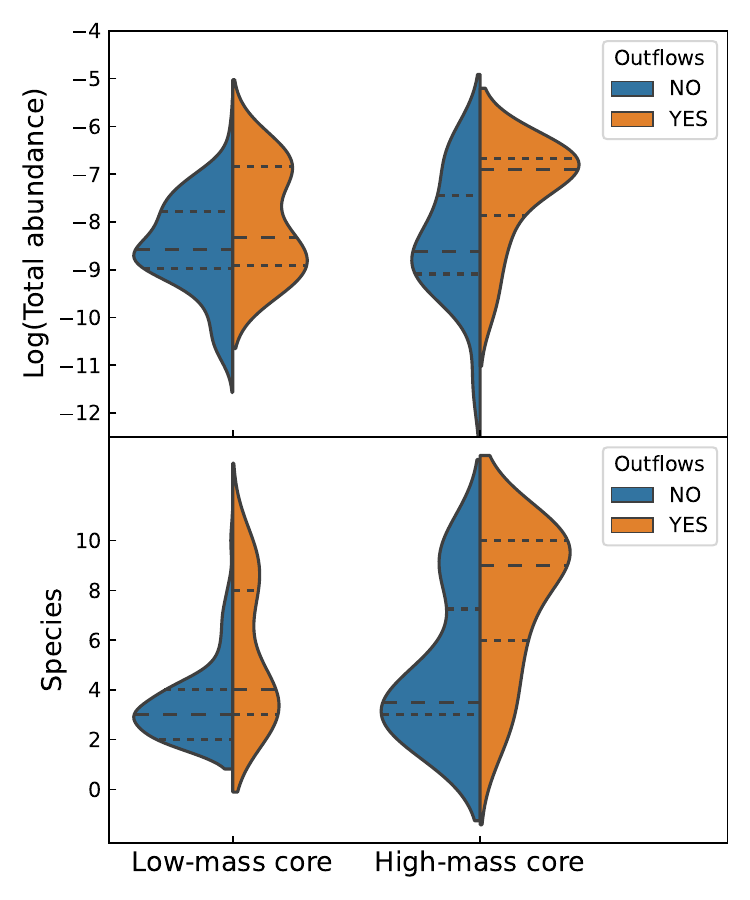}
	\caption{Violin plots for total abundance (upper panel) and the number of detected species (bottom panel). The regions with orange and blue colors represent the presence and absence of outflows, respectively. In each panel, the left plots are statistics of low-mass cores, and the right plots are statistics of high-mass cores. The dotted lines in the plots represent the quartiles.} 
	\label{fig:violinplot_outflow}
\end{figure}

It is well-established that outflows can enahnce S-bearing species in the gas phase \cite[e.g.][]{Pineau1993MNRAS.262..915P,Charnley1997ApJ...481..396C, Hatchell1998A&A...338..713H,Viti2001A&A...370.1017V,Taquet2020A&A...637A..63T,Artur2023A&A...678A.124A,Zhang2023ApJ...946..113Z}. Based on the information provided by \cite{Xu2024ApJS..270....9X}, we categorized the protostellar cores into two groups: the presence and absence of outflows. Comparisons of abundances of each species were conducted. We only found that the total abundances and the number of detected species in high-mass cores with outflows exceeded those cores without outflows. But this trend was less pronounced in low-mass cores. A comparison of the total abundance and the number of detected S-bearing species between these two categories is presented in Figure \ref{fig:violinplot_outflow}. 
Our result suggests that outflows have indeed elevated the abundances and detection rates of S-bearing species in high-mass cores, underscoring the significant impact of high-mass star-forming feedback on the formation of S-bearing species. This resuls align with our earlier discussion in Section \ref{sec:low&high}.
	
\subsection{Sulfur depletion}\label{sec:depletion}
\begin{figure}
	\plotone{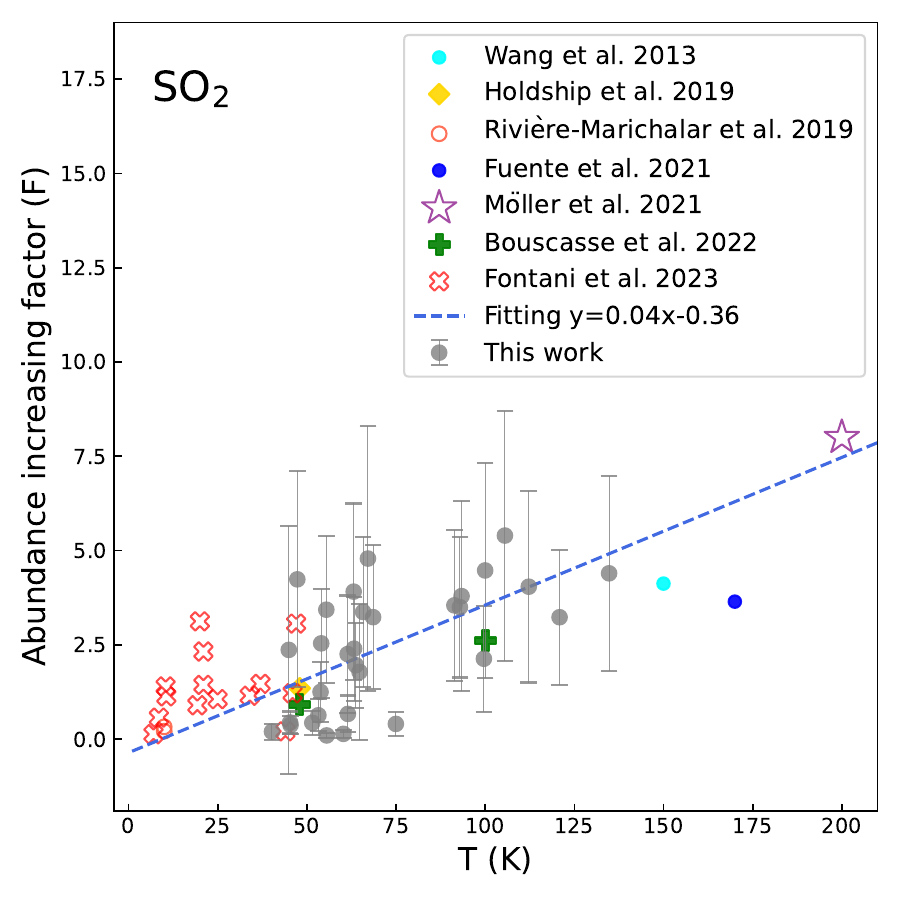}
	\caption{Relationship between the abundance-increasing factor $F$ of \sot~and temperatures. The linear fitting is demonstrated as the blue dashed line. The fitting results are shown in legend. The definition of restoration factor can be found in Section \ref{sec:depletion}. The $F$ of other studies are estimated by the abundances and gas temperatures collected from literatures \cite[][]{Wang2013A&A...558A..69W,Holdship2019ApJ...878...64H,Rivi2019A&A...628A..16R,Fuente2021MNRAS.507.1886F,Moller2021A&A...651A...9M,Bouscasse2022A&A...662A..32B,Fontani2023A&A...680A..58F}.} 
	\label{fig:AIF}
\end{figure}

As stated in Section \ref{sec:introduction}, the issue of sulfur depletion has been observed over the past few decades, but the reservoir of the sulfur element is still unknown. 
Our observed total S-bearing molecular abundance ranges from 4.4$\times$10$^{-12}$ to 6.9$\times$10$^{-7}$. 
All of these values are lower than the sulfur abundance in the diffuse and highly ionized regions of $\sim$10$^{-5}$ \cite[e.g.][]{Savage1996ApJ...470..893S,2002A&A...381..606M,How2006ApJ...637..333H,2006MNRAS.368..253G,Yamamoto2017iace.book.....Y}, indicating sulfur depletion is occurring in all dense cores.
Possible solid sulfur reservoir candidates, for instance, H$_{2}$S, OCS, \sot, organo-sulfur species, sulfur chains (S$_{n}$, $n$=2$\cdots$8), and sulfide minerals (FeS and MgS) were proposed \cite[][]{Keller2002Natur.417..148K,Hony2002A&A...393L.103H,Scappini2003MNRAS.341..657S,Wakelam2004A&A...422..159W,Jim2011A&A...536A..91J,Druard2012MNRAS.426..354D,Vidal2017MNRAS.469..435V,Laas2019A&A...624A.108L,Kama2019ApJ...885..114K}. 
Unfortunately, only OCS  \cite[]{Geballe1985A&A...146L...6G,Palumbo1995ApJ...449..674P,McClure2023NatAs...7..431M} and \sot~ \cite[]{Boogert1997A&A...317..929B,Yang2022ApJ...941L..13Y,McClure2023NatAs...7..431M, Rocha2024A&A...683A.124R} have been detected in interstellar ices.

The solid sulfur reservoir may sublimate from grains to gas phase as the temperature increases, making it a temperature-sensitive species. The temperature tracers \htcs~and \tfsot~enable us to determine the gas temperatures of S-bearing species in dense cores. A positive correlation between molecular abundance and gas temperature maybe a indicator of a sulfur reservoir.
However, as temperatures elevate, chemical reactions also intensify, potentially leading to increased gas-phase abundances of numerous species. Therefore, assessing the correlation between abundance and gas temperature alone may be inadequate to investigate sulfur depletion. This is because many species, not just sulfur-bearing ones, could exhibit positive correlations with temperature, complicating the interpretation of the depletion process.

Solid sulfur reservoirs, such as OCS and \sot, are anticipated to exhibit heightened temperature sensitivity compared to other S-bearing species due to they are locked onto ice grains. With the ascent of temperature, the abundances of these species are predicted to escalate at a more accelerated pace relative to other S-bearing species. This is attributed to their dual production pathways, which encompass not only gas-phase reactions but also the desorption mechanism from dust grains into the gas phase.
In light of this, we have introduced the abundance increasing factor ($F$) to quantitatively evaluate the rate at which the abundance of a specific sulfur-bearing species increases in comparison to other S-bearing species.

\begin{equation}
	F= \frac{f_{\rm x}}{(f_{\rm total}/n)},
\end{equation}
where $f_{\rm x}$ represents the abundance of a certain S-bearing species, $f_{\rm total}$ denotes the total S-bearing molecular abundance, and $n$ is the number of detected S-bearing species in a dense core.

Consequently, we have calculated the abundance increasing factor ($F$) for \sot~exclusively within dense cores. A value of $F$ exceeding unity for \sot~suggests that it is more abundant than other S-bearing species in a given dense core. An elevated $F$ value denotes a relatively higher abundance of \sot~in comparison to other S-bearing species. Thereby, the correlation between $F$ and the \tone\footnote{We adoptted \tone~ instead of \ttwo~to investigate the correlations for three reasons: (1) \ttwo~primarily traces high-temperature central regions where are not the main regions of sulfur-depletion; (2) \ttwo~is unsuitable for the correlation analysis due to \sot~abundance and \ttwo~are interdependent parameters; (3) Multiple \htcs~line transitions are observed in most of protostellar cores, and then \htcs~can be used as a temperature tracer in various physical environments.} serves as a metric for gauging the rate at which \sot~abundance increases. The interrelation between $F$ and gas temperature for \sot~is illustrated in Figure \ref{fig:AIF}. 
A linear correlation has been found with a slope of 0.04 and an intercept of --0.36.
For comparison, we have collated abundances and gas temperatures from analogous studies to estimate $F$, and these literature-derived results are concurrently presented in the figure.

A linear regression analysis was conducted on the samples within our study, which disclosed a positive correlation between the abundance increasing factor ($F$) and temperature. This result is corroborated by the results of previous studies, indicating a consistency in the observed correlation.
The presence of this correlation suggests that \sot~is likely to be significantly depleted onto grain mantels and may undergo desorption as the temperature elevates. This desorption process could accelerate the increase in \sot~abundance in the gas phase relative to other sulfur-bearing species.
Furthermore, this correlation potentially explains the relatively high abundance of \sot~($\sim$10$^{-7}$) observed in star-forming regions. \sot~can be produced through both gas-phase reactions and direct desorption from grain mantles.
Additionally, as shown in Figure \ref{fig:cores}, the \sot~emission is more compact and probably coming from the desorption of ice mantles located at higher temperatures, closer to the protostar.

However, our observations indicate that the abundance of \sot~constitutes only approximately 1\% of the total sulfur inventory. When accounting for other volatile species, such as H$_{2}$S and OCS, their combined contribution remains minimal, at less than 4\% \cite[][]{Geballe1985A&A...146L...6G,Palumbo1995ApJ...449..674P,Boogert1997A&A...317..929B,Kama2019ApJ...885..114K}.
In areas influenced by shocks, the sulfur abundance may be elevated to approximately 10\% of the total sulfur inventory \cite[][]{Anderson2013ApJ...779..141A}, yet approximately 90\% of the sulfur remains unaccounted for. Our results, combined with analogous studies, imply that while \sot~could serve as a sulfur reservoir on the grain mantles of dense cores, but the predominant solid sulfur reservoir is more likely to be composed of refractory elements. For instance, the sulfide mineral FeS, as proposed by \cite{Hony2002A&A...393L.103H,Kama2019ApJ...885..114K}, is a strong candidate due to its high sublimation temperature of around 655 K \cite[][]{Larimer1967GeCoA..31.1215L,Lodders2003ApJ...591.1220L,Kama2019ApJ...885..114K}.

\section{Summary} \label{sec:summary}
Based on the ALMA Band 7 (870 $\micron$) data, we have investigated the properties of S-bearing molecules in 248 dense cores from 11 massive star-forming clumps.
The main results of this study are summarized as follows:
\begin{enumerate}
		\item Ten S-bearing species of CS, SO, \htcs, NS, \sot, \ttso, \octfs, \tfsot, \ttsot, and \seoo~are detected. The parameters of them (except CS and NS) were estimated and presented.
                 
        \item Our analysis shows that \htcs~and SO exhibit similar gas distribution. The 2D KDE analysis suggests they have comparable abundance properties, with levels of $\sim$10$^{-9}$ in prestellar core candidates and $\sim$10$^{-8}$ in protostellar cores. SO appears to be more easily detected in a broader range of physical and chemical environments than \htcs~despite many similarities found between SO and \htcs.
        
        \item Temperatures were determined from multiple transitions of \htcs~and \tfsot. Gas distributions and 2D KDE analyses indicate that \htcs~and \tfsot~trace regions with varying temperatures. \htcs~appears to trace the outer warm envelope, while \tfsot~is associated with high-temperature central-regions.
      
        \item Abundances of S-bearing species and detection rates are higher in high-mass cores. SO and \sot~show correlations between abundances and $\sigma_{NT}$ in low-mass cores, suggesting non-thermal motions may enhance these species. Particularly for \sot, its abundance and non-thermal velocity dispersion ($\sigma_{NT}$) appear notably increased in high-mass cores, suggesting that feedback from high-mass star formation significantly boosts the abundance of \sot.
       
        \item  The outflows elevate the total abundance and detection rates of S-bearing species in high-mass cores, underscoring the significant impact of high-mass star-forming feedback on the formation of S-bearing species.
        
		\item  Sulfur depletion was observed to be prevalent in dense cores. A positive correlation between abundance increasing factor ($F$) and temperature for \sot~was discovered. The results from similar previous works also conform to this correlation, suggesting that a large amount of \sot~may exist in the solid phase and desorb rapidly into a gaseous phase as the temperature rises. This result can also explain a high \sot~abundance of $\sim$10$^{-7}$ and compact gas distributions in massive star-forming regions. However, \sot~abundance only accounts for $\sim$1\% of the total sulfur inventory, implying that the predominant solid sulfur reservoirs would be more likely to be composed of refractory species with higher sublimation temperatures.
\end{enumerate}

In conclusion, our study presents statistical findings on S-bearing species across 249 dense cores, including low-mass, high-mass, prestellar, and protostellar cores. The analysis, based on a large sample size, offers valuable insights of sulfur-depletion and properties of S-bearing species. However, the observed number of S-bearing species remains limited. Future investigations on S-bearing species could be enhanced by expanding observations to include as many as unblending transitions in larger sample sizes, leading to more accurate parameters estimations.

\begin{acknowledgments}
The authors deeply appreciate the crucial comments from the anonymous reviewers, which have undoubtedly improved the quality of this paper.
This work has been supported by the National Science Foundation of China (12203011, 12033005). 
MYT acknowledge the support by Yunnan province talent support program, Yunnan provincial Department of Science and Technology through grant No.202101BA070001-261, Yunnan University Laboratory Open Project, and PhD research startup foundation of Chuxiong Normal University.
S.-L.Q. thanks the Xinjiang Uygur Autonomous Region of China for their support through the Tianchi Program.
T.L by the international partnership program of Chinese Academy of Sciences through grant No.114231KYSB20200009, Shanghai Pujiang Program 20PJ1415500, the science research grants from the China Manned Space Project with no. CMS-CSST-2021-B06.
L.A.Z. acknowledges financial support from CONACyT-280775, UNAM-PAPIIT IN110618, and IN112323 grants, M\'{e}xico.
Y.P. Peng acknowledges support from NSFC through grant No. 12303028.

This paper uses the following ALMA data: ADS/JAO.ALMA\#2017.1.00545.S.
ALMA is a partnership of ESO (representing its member states), NSF (USA) and NINS (Japan), together with NRC (Canada), MOST and ASIAA (Taiwan), and KASI (Republic of Korea), in cooperation with the Republic of Chile. 
The Joint ALMA Observatory is operated by ESO, AUI/NRAO, and NAOJ. 
\end{acknowledgments}

\vspace{5mm}
\facilities{ALMA(Atacama Large Milimeter/Submilimeter Array)}

\software{astropy \cite[][]{2013A&A...558A..33A,2018AJ....156..123A,2022ApJ...935..167A},
          MADCUBA\cite[][]{2019A&A...631A.159M}, CASA\cite[][]{2022PASP..134k4501C}
          }

\appendix
\section{Self-calibration}\label{Sec:selfcal}
\setcounter{table}{0}
\renewcommand{\thetable}{A\arabic{table}}
\begin{deluxetable}{cccccccc} 
	\label{tab:selfcal}
	\tabletypesize{\tiny}
	\tablecaption{Improvements of Signal-to-noise ratios (SNRs) of self-calibrations}
	\tablewidth{0pt} \tablehead{
		\colhead{Source name} &\colhead{SNR$^{a}$} &SNR$^{b}$  &SNR$^{c}$  &Improvement$^{d}$ \\
		\colhead{}            &\colhead{(no selfcal)} &(phase-only)   &\colhead{(phase \& amp)}      &                                
	}
	\startdata
IRAS 14382-6017 &146 &190  &198 &36\% \\
IRAS 14498-5856 &372 &446  &501 &35\% \\
IRAS 15520-5234 &201 &295  &296 &47\% \\
IRAS 15596-5301 &273 &336  &353 &29\% \\
IRAS 16060-5146 &489 &605  &616 &26\% \\
IRAS 16071-5142 &566 &647  &649 &15\% \\
IRAS 16076-5134 &300 &354  &360 &20\% \\
IRAS 16272-4837 &759 &838  &838 &10\% \\
IRAS 16351-4722 &340 &353  &370 &9\% \\
IRAS 17204-3636 &224 &270  &279 &25\% \\
IRAS 17220-3609 &266 &280  &281 &5\% \\
\enddata
	\tablecomments{\\
	    $^{a}$: The SNR of without performing self-calibration. \\
		$^{b}$: The SNR after performing phase-only self-calibration. \\
	    $^{c}$: The SNR after performing phase-only and amplitude self-calibration. \\
	    $^{d}$: The improvement of SNR by comparing images before and after self-calibration. }
	\end{deluxetable}
	
Self-calibration could be a crucial technique for enhancing image quality when the signal-to-noise ratio (SNR) exceeds 20 \cite[][]{Brogan2018arXiv180505266B}. As shown in Table \ref{tab:selfcal}, our data quality justifies the use of self-calibration. We conducted three rounds of phase-only self-calibration followed by one round of amplitude self-calibration.

The phase-only self-calibration began with a solution interval of a scan length (\texttt{solint=inf}), which is ideal for the initial phase-only self-calibration \cite[][]{Richards2022arXiv220705591R}. The second round was set with a solution interval of 6.05s (\texttt{solint=int}). If this setting results in over 10\% failed solutions, intervals of 12.1s or 18.15s will be used. The third round of phase-only self-calibration utilized 6.05s to properly correct the phase over time. Table \ref{tab:selfcal} presents the self-calibration results for each source, showing that the phase-only self-calibrations improved the SNRs of the images.

Phase-only self-calibrations can alter image fluxes, prompting us to conduct a subsequent round of amplitude self-calibration. The amplitude self-calibration has a greater potential to affect images than phase-only self-calibration. To avoid systemic errors, we set the solution interval to \texttt{solint=inf} since amplitude does not vary rapidly over time. We also enabled \texttt{solnorm=True} to maintain the ensemble amplitude correction around 1, which helps prevent an overall reduction in integrated flux \cite[][]{Brogan2018arXiv180505266B}. Finally, we applied the amplitude solutions alongside the best phase solutions. The results in Table \ref{tab:selfcal} indicate that amplitude self-calibrations did not induce systemic errors in the images.  The SNRs improved by 5\% to 47\% compared to images without self-calibration.

\section{Parameters derived from spectra}\label{sec:A}
In this section, we present the spectroscopic parameters of detected species in Table \ref{tab:transitions}. The parameters of systemic velocities, line widths, temperatures, and column densities of detected S-bearing species in each core are presented in Table \ref{tab:h2cs} to Table \ref{tab:33so2}. The dense cores without detection of S-bearing species are excluded from the tables.

The derivation of these parameters was executed by \texttt{SLIM} from the \texttt{MADCUBA} package \cite[][]{2019A&A...631A.159M}. \texttt{SLIM} is capable of generating synthetic spectra under the LTE assumption, utilizing five input parameters: excitation temperature, column density, line width, velocity, and source size. The integrated ``auto-fit'' feature within \texttt{SLIM} facilitates the best non-linear least-squares fit, employing the Levenberg-Marquardt algorithm to fit the LTE model on the observed spectra. The output of the fitting process encompasses parameters such as column density, temperature, velocity, line width, and source size.
    
The spectral data for each core were extracted from pixels that coincide with the peak positions of the continuum flux. Continuum intensity has been subtracted from these spectra. The resultant reduced spectra were then imported into \texttt{SLIM} as observational data for the fitting procedure. During this process, the source size for each core was set as a fixed parameter, with the core sizes referenced from \cite{Xu2024ApJS..270....9X}. For molecules \htcs~and \tfsot, the column density, temperature, velocity, and line width were treated as free parameters during the fitting. For \ttso, \octfs, \sot, \ttsot, and \seoo, the fitting was performed with column density, velocity, and line width as the free parameters.    

\setcounter{table}{0}
\renewcommand{\thetable}{B\arabic{table}}

\begin{deluxetable*}{cccccccccc} 
	\label{tab:transitions}
	\tabletypesize{\tiny}
	\tablecaption{Detected transitions and corresponding spectroscopic parameters}
	\tablewidth{0pt} \tablehead{
		\colhead{Molecule}       &\colhead{Resolved QNs}  &\colhead{Rest frequency}                  &\colhead{Upper energy level}   &\colhead{S$_{\rm ij}\mu^2$}    &log$_{10}$(A$_{ij}$)  &Comment$^{(a)}$ \\
		\colhead{}                   &\colhead{}                          &\colhead{MHz}                                      &\colhead{(K)}                              &\colhead{$D^2$}                                   &\colhead{}     &
	}
	\startdata
	CS            &7-6                      &342882.85 &65.83     &26.75  &-3.08 &$^{\times}$ \\
	\hline
	H$_{2}$CS              &10(0,10)-9(0,9) &342946.42 &90.59      &27.20  &	-3.22 &$\ast$ \\
	&10(6,5)-9(6,4)   &343081.52  &563.28    &17.40  &-3.41 \\
	&10(5,6)-9(5,5)   &343203.24 &419.17    &61.20   &-3.34 \\
	&10(4,7)-9(4,6)   &343309.83 &301.07    &22.84   &-3.29 \\
	&10(2,9)-9(2,8)   &343322.08 &143.31    &26.11  &-3.23 \\
	&10(3,8)-9(3,7)   &343409.96 &209.09  &74.24  &-3.26 &$\ast$\\
	&10(3,7)-9(3,6)   &343414.15  &209.09   &74.24  &-3.26 &$\ast$ \\
\enddata
	{$^{(a)}$: \\
		$\times$: Transitions of CS and NS molecules are detected. The CS line is typically optically thick, displaying distinct absorption profiles and outflow wings. NS lines are heavily impacted by line blending; \\
		$\ast$: The transitions adopted to estimate gas temperatures; \\
		Blended: The transition is blended with other molecular transitons, and excluded from the presentation in Figure \ref{fig:spectra}.}
	\tablecomments{The basic information of transitions are from CDMS\citep{CDMS}.}
\end{deluxetable*}

\clearpage
\startlongtable
\begin{deluxetable*}{cccc|cc|cc|cccccc}
	\tablecaption{\htcs~parameters.\label{tab:h2cs}}
	\setlength{\tabcolsep}{0.03in}
	\tablewidth{0pt}
	\tabletypesize{\tiny}
	\tablehead{
	\colhead{Core} &Comps$^{a}$ &\multicolumn{2}{c}{Parameters} &\multicolumn{2}{c}{Component1}   &\multicolumn{2}{c}{Component2} &\multicolumn{2}{c}{Component3} \\
	\cline{3-10}
	&  &$T$  &$N_{total}^{b}$ &$V_{\rm lsr}$ &FWHM &$V_{\rm lsr}$ &FWHM &$V_{\rm lsr}$ &FWHM \\
	& &(K)   &(cm$^{-2}$) &(km s$^{-1}$) &(km s$^{-1}$) &(km s$^{-1}$) &(km s$^{-1}$)  &(km s$^{-1}$) &(km s$^{-1}$) 
}
\startdata
I14382-8	&1	&51.7(10.7)  &8.3(1.9)E+13 	&-60.0(0.2)    &2.4(0.5) 	&					& & & & & \\
I14382-11	&1	&50.3(7.7)   &1.7(0.3)E+14 	&-59.2(0.1)    &2.4(0.3) 	&					& & & & & \\
I14382-12	&1	&45.3(7.3)   &6.0(1.0)E+13 	&-59.3(0.1)    &1.4(0.2) 	&					& & & & & \\
I14382-14	&1	&60.8(9.9)   &7.8(1.2)E+13 	&-59.9(0.2)    &3.7(0.6) 	&					& & & & & \\
I14382-15	&1	&28.3(9.2)   &9.5(1.6)E+13 	&-60.1(0.2)    &1.9(0.4) 	&					& & & & & \\
I14498-5	&1	&73.7(7.7)   &7.6(0.8)E+14 	&-52.2(0.1)    &5.6(0.2) 	&					& & & & & \\																						
I14498-7	&1	&105.5(11.0) &9.5(1.0)E+15 	&-50.9(0.1)    &6.6(0.1) 	&					& & & & & \\
I14498-8	&1	&68.7(8.0)   &4.3(0.5)E+14 	&-49.8(0.1)    &3.1(0.1) 	&					& & & & & \\
\enddata
	\tablecomments{\\
		$a$: The number of velocity components defined in the spectra.\\ 
	    $b$: The column density are calculated by intergrating all different components.}
\end{deluxetable*}

\startlongtable
\begin{deluxetable*}{cccc|cc|cc}
	\tablecaption{\tfsot~parameters.\label{tab:34so2}}
	\setlength{\tabcolsep}{0.03in}
	\tablewidth{0pt}
	\tabletypesize{\tiny}
	\tablehead{
		\colhead{Core} &Comps$^{a}$ &\multicolumn{2}{c}{Parameters}  &\multicolumn{2}{c}{Component1}  &\multicolumn{2}{c}{Component2}\\
		\cline{3-8}
		&  &$T$  &$N_{total}^{b}$ &$V_{\rm lsr}$ &FWHM  &$V_{\rm lsr}$ &FWHM  \\
		& &(K)   &(cm$^{-2}$) &(km s$^{-1}$) &(km s$^{-1}$)  &(km s$^{-1}$) &(km s$^{-1}$) 
	}
	\startdata
I14498-7	&1	&121.6(13.2)   &8.1(0.8)E+15	&-50.8(0.1)	 &7.8(0.1)	&  & \\							
I15520-18	&2	&105.3(11.0)   &5.1(1.1)E+15	&-41.0(0.4)	 &4.6(0.7)	&-36.6(0.3)	&4.0(0.5) \\
I15520-21	&1	&158.0(127.2)  &2.8(1.8)E+14	&-44.3(0.1)	 &2.0(0.5)	&  & \\							
I15520-22	&1	&116.9(19.0)   &5.5(0.7)E+15	&-42.9(0.2)	 &6.3(0.4)	&  & \\							
I15596-14	&1	&97.6(22.5)    &6.9(1.0)E+14	&-71.9(0.3)	 &5.9(0.5)	&  & \\							
I15596-19	&1	&99.1(12.9)    &3.7(0.4)E+15	&-71.4(0.2)	 &8.1(0.4)	&  & \\							
I16060-2	&1	&104.0(11.6)   &6.9(0.7)E+14	&-89.8(0.1)	 &3.5(0.1)	&  & \\							
I16060-4	&1	&112.8(12.3)   &2.3(0.2)E+16	&-94.3(0.1)	 &4.3(0.1)	&  & \\
\enddata
	\tablecomments{\\
	$a$: The number of velocity components defined in the spectra.\\ 
	$b$: The column density are calculated by intergrating all different components.}
\end{deluxetable*}

\startlongtable
\begin{deluxetable*}{cccc|cc|cc|cc|cc}
	\tablecaption{SO parameters.\label{tab:so}}
	\setlength{\tabcolsep}{0.03in}
	\tablewidth{0pt}
	\tabletypesize{\tiny}
	\tablehead{
		\colhead{Core} &Comps$^{a}$ &\multicolumn{2}{c}{Parameters}  &\multicolumn{2}{c}{Component1} &\multicolumn{2}{c}{Component2} &\multicolumn{2}{c}{Component3} &\multicolumn{2}{c}{Component4}\\
		\cline{3-12}
		&  &$T$  &$N_{total}^{b}$ &$V_{\rm lsr}$ &FWHM &$V_{\rm lsr}$ &FWHM &$V_{\rm lsr}$ &FWHM &$V_{\rm lsr}$ &FWHM \\
		& &(K)   &(cm$^{-2}$) &(km s$^{-1}$) &(km s$^{-1}$) &(km s$^{-1}$) &(km s$^{-1}$) &(km s$^{-1}$) &(km s$^{-1}$) &(km s$^{-1}$) &(km s$^{-1}$
	}
	\startdata
I14382-2	&1	&70.7(35.8)  &5.4(0.7)E+13		&-61.0(0.1) 	&1.4(0.1) 	&				&				\\								
I14382-4	&2	&28.0(2.8)	 &1.2(0.2)E+14		&-57.1(0.2) 	&1.2(0.4) 	&-60.2(0.2) 	&2.1(0.4) 		\\														
I14382-8	&1	&51.7(10.7)  &7.4(1.0)E+13		&-60.2(0.1) 	&2.2(0.2) 	&				&	            \\
I14382-11	&1	&50.3(7.7)	 &3.0(0.3)E+14		&-59.3(0.3) 	&2.0(0.1) 	&				&	            \\
I14382-12	&1	&45.3(7.3)	 &9.6(1.0)E+13		&-59.6(0.1) 	&1.6(0.1) 	&				&	            \\
I14382-14	&1	&60.8(9.9)	 &1.2(0.1)E+14		&-60.1(0.1) 	&3.2(0.2) 	&				&	            \\
I14382-15	&1	&28.3(9.2)	 &5.8(0.8)E+13		&-60.3(0.1) 	&1.1(0.3) 	&				&	            \\
I14382-16	&1	&28.0(2.8)	 &3.2(0.8)E+13		&-59.1(0.2) 	&1.3(0.3) 	&				&	            \\
\enddata
	\tablecomments{\\
		$a$: The number of velocity components defined in the spectra. \\
		$b$: The column density are calculated by intergrating all different components.\\
		absorption: The component is observed as absorption. Their parameters of $V_{lsr}$ and FHWM are not estimated. The column densities are calculated by equation (\ref{eq:1}).}
\end{deluxetable*}

\startlongtable
\begin{deluxetable*}{cccc|cc|cc|cc|cc}
	\tablecaption{\ttso~parameters.\label{tab:33so}}
	\setlength{\tabcolsep}{0.03in}
	\tablewidth{0pt}
	\tabletypesize{\tiny}
	\tablehead{
		\colhead{Core} &Comps$^{a}$ &\multicolumn{2}{c}{Parameters}  &\multicolumn{2}{c}{Component1} &\multicolumn{2}{c}{Component2} &\multicolumn{2}{c}{Component3} &\multicolumn{2}{c}{Component4}\\
		\cline{3-12}
		&  &$T$  &$N_{total}^{b}$ &$V_{\rm lsr}$ &FWHM &$V_{\rm lsr}$ &FWHM &$V_{\rm lsr}$ &FWHM &$V_{\rm lsr}$ &FWHM\\
		& &(K)   &(cm$^{-2}$) &(km s$^{-1}$) &(km s$^{-1}$) &(km s$^{-1}$) &(km s$^{-1}$) &(km s$^{-1}$) &(km s$^{-1}$) &(km s$^{-1}$) &(km s$^{-1}$) 
	}
	\startdata
I14498-7	&2	&105.5(11.0)  &1.9(0.2)E+15 	&-50.1(0.1) 	&8.5(0.2) 	&-42.8(0.2) 	&4.2(0.4) 	&               &			\\						
I15520-16	&1	&60.2(6.2)    &2.4(0.4)E+13 	&-42.9(0.1) 	&1.3(0.4) 	&				&			&               &           \\
I15520-18	&1	&71.0(8.0)    &1.2(0.1)E+15 	&-39.9(5.0) 	&7.2(0.1) 	&				&			&               &           \\
I15520-19	&1	&68.1(5.4)    &9.1(1.0)E+13 	&-40.4(0.1) 	&3.2(0.2) 	&				&			&               &           \\
I15520-21	&2	&80.8(10.3)   &3.0(0.4)E+14 	&-44.6(0.1) 	&1.9(0.2) 	&-39.0(0.3) 	&2.5(0.9) 	&               &           \\
I15520-22	&1	&83.2(6.3)    &1.7(0.2)E+15 	&-42.4(0.1) 	&6.1(0.1) 	&				&			&               &           \\
I15596-14	&1	&81.0(7.8)    &5.4(0.5)E+14 	&-70.6(0.1) 	&0.5(0.2) 	&				&			&               &           \\
I15596-19	&1	&93.4(9.6)    &1.0(0.1)E+15 	&-70.4(0.1) 	&10.5(0.2) 	&				&			&               &           \\
\enddata
	\tablecomments{\\
		$a$: The number of velocity components defined in the spectra.\\
		$b$: The column density are calculated by intergrating all different components.}
\end{deluxetable*}

\startlongtable
\begin{deluxetable*}{cccc|cc|cc|cc}
	\tablecaption{\octfs~parameters.\label{tab:oc34s}}
	\setlength{\tabcolsep}{0.03in}
	\tablewidth{0pt}
	\tabletypesize{\tiny}
	\tablehead{
		\colhead{Core} &Comps$^{a}$ &\multicolumn{2}{c}{Parameters}  &\multicolumn{2}{c}{Component1} &\multicolumn{2}{c}{Component2} &\multicolumn{2}{c}{Component3}\\
		\cline{3-10}
		&  &$T$  &$N_{total}^{b}$ &$V_{\rm lsr}$ &FWHM &$V_{\rm lsr}$ &FWHM  &$V_{\rm lsr}$ &FWHM \\
		& &(K)   &(cm$^{-2}$) &(km s$^{-1}$) &(km s$^{-1}$)  &(km s$^{-1}$) &(km s$^{-1}$) &(km s$^{-1}$) &(km s$^{-1}$) 
	}
	\startdata
I14498-7	&1	&105.5(11.0)   &6.9(0.7)E+15  &-50.6(0.1) 	&6.0(0.2) 	&				&			& & \\ 
I15520-15	&1	&83.4(9.1)     &9.5(1.6)E+14  &-40.5(0.2) 	&3.8(0.6) 	&				&			& & \\ 
I15520-18	&1	&71.0(8.0)     &3.3(0.3)E+15  &-40.9(0.1) 	&5.3(0.1) 	&				&			& & \\ 
I15520-19	&1	&68.1(5.4)     &1.1(0.1)E+15  &-40.6(0.1) 	&3.1(0.2) 	&				&			& & \\ 
I15520-20	&1	&63.2(5.2)     &7.8(0.9)E+14  &-42.8(0.1) 	&1.8(0.1) 	&				&			& & \\ 
I15520-21	&1	&80.8(10.3)    &6.0(0.6)E+15  &-44.2(0.1) 	&2.0(0.1) 	&				&			& & \\ 
I15520-22	&1	&83.2(6.3)     &9.3(1.0)E+15  &-43.5(0.1) 	&3.6(0.2) 	&				&			& & \\ 
I15596-14	&1	&81.0(7.8)     &1.9(0.2)E+15  &-71.3(0.2) 	&6.3(0.2) 	&				&			& & \\
\enddata
	\tablecomments{\\
	$a$: The number of velocity components defined in the spectra.\\
	$b$: The column density are calculated by intergrating all different components.}
\end{deluxetable*}

\startlongtable
\begin{deluxetable*}{cccc|cc|cc}
	\tablecaption{\sot~parameters.\label{tab:so2}}
	\setlength{\tabcolsep}{0.03in}
	\tablewidth{0pt}
	\tabletypesize{\tiny}
	\tablehead{
		\colhead{Core} &Comps$^{a}$ &\multicolumn{2}{c}{Parameters}  &\multicolumn{2}{c}{Component1} &\multicolumn{2}{c}{Component2}\\
		\cline{3-8}
		&  &$T$  &$N_{total}^{b}$ &$V_{\rm lsr}$ &FWHM &$V_{\rm lsr}$ &FWHM  \\
		& &(K)   &(cm$^{-2}$) &(km s$^{-1}$) &(km s$^{-1}$) &(km s$^{-1}$) &(km s$^{-1}$) 
	}
	\startdata
I14498-7	&1	&121.6(13.2) 	&9.9(2.5)E+16	&-50.8(0.2) 	&7.8(0.3) 	&				&				\\						
I15520-4	&1	&60.7(7.5)	    &3.0(0.3)E+14	&-43.9(0.1) 	&3.7(0.3) 	&				&				\\						
I15520-15	&1	&68.2(7.3)	    &2.5(0.5)E+14	&-39.0(0.2) 	&2.3(0.4) 	&				&				\\						
I15520-16	&1	&60.2(6.2)	    &1.7(0.7)E+14	&-42.7(0.1) 	&1.1(0.1) 	&				&				\\						
I15520-18	&2	&106.0(15.5) 	&3.3(1.0)E+16	&-41.3(0.3) 	&4.7(0.7)	&-36.8(0.7) 	&3.9(0.9) 		\\													
I15520-19	&1	&70.1(7.6)	    &7.2(1.4)E+14	&-40.2(0.2) 	&3.0(0.6) 	&				&				\\						
I15520-21	&1	&158.0(127.2) 	&5.6(0.6)E+15	&-44.4(0.1) 	&2.2(0.1) 	&				&				\\						
I15520-22	&1	&116.9(19.0) 	&4.0(0.4)E+16	&-43.1(0.1) 	&5.7(0.2) 	&				&				\\
\enddata
	\tablecomments{\\
	$a$: The number of velocity components defined in the spectra.\\
	$b$: The column density are calculated by intergrating all different components.}
\end{deluxetable*}

\startlongtable
\begin{deluxetable*}{cccc|cc}
	\tablecaption{\seoo~parameters.\label{tab:so18o}}
	\tablewidth{700pt}
	\tabletypesize{\tiny}
	\tablehead{
		\colhead{Core} &Comps$^{a}$ &\multicolumn{2}{c}{Parameters} &\multicolumn{2}{c}{Component1}\\
		\cline{3-6}
		&  &$T$  &$N_{total}$ &$V_{\rm lsr}$ &FWHM  \\
		& &(K)   &(cm$^{-2}$) &(km s$^{-1}$) &(km s$^{-1}$)
	}
	\startdata
I14498-7	&1	&121.62(13.2)	&9.9(3.3)E+14	&-50.9(0.1) 	&6.2(0.1) 	\\								
I15520-18	&1	&104.6(15.5)	&6.6(0.9)E+14	&-37.7(0.2) 	&5.1(0.5) 	\\								
I15520-21	&1	&158.0(127.2) 	&1.6(0.4)E+14	&-40.2(0.2) 	&1.7(0.4) 	\\								
I16060-4	&1	&112.8(12.1)	&4.1(0.4)E+15	&-94.2(0.1) 	&4.1(0.2) 	\\								
I16060-7	&1	&100.7(11.2)	&1.1(0.3)E+16	&-87.5(0.6) 	&4.6(1.4) 	\\								
I16060-8	&1	&122.6(12.9)	&1.8(0.2)E+16	&-94.5(0.1) 	&7.0(0.2) 	\\								
I16060-9	&1	&112.2(12.4)	&3.1(0.3)E+16	&-94.3(0.1) 	&4.8(0.1) 	\\								
I16060-10	&1	&119.8(15.5)	&8.7(2.0)E+15	&-88.2(0.7) 	&5.7(0.1) 	\\								
I16060-11	&1	&113.2(13.7)	&1.0(0.1)E+16	&-90.8(0.3) 	&6.2(0.1) 	\\								
I16076-15	&1	&118.4(13.5)	&3.6(0.5)E+15	&-91.8(0.5) 	&8.0(0.1) 	\\								
I16272-2	&1	&142.5(41.6)	&1.7(0.2)E+15	&-45.0(0.1) 	&6.2(0.1) 	\\								
I17220-17	&1	&102.0(11.3)	&4.9(1.3)E+15	&-93.3(0.1) 	&7.3(0.1) 	\\								
I17220-18	&1	&104.4(11.4)	&9.8(6.5)E+15	&-94.8(0.1) 	&4.0(0.1) 	\\													
	\enddata
	\tablecomments{\\
	$a$: The number of velocity components defined in the spectra.}
\end{deluxetable*}

\startlongtable
\begin{deluxetable*}{cccc|cc|cc}
	\tablecaption{\ttsot~parameters.\label{tab:33so2}}
	\setlength{\tabcolsep}{0.03in}
	\tablewidth{0pt}
	\tabletypesize{\tiny}
	\tablehead{
		\colhead{Core} &Comps$^{a}$ &\multicolumn{2}{c}{Parameters}  &\multicolumn{2}{c}{Component1}  &\multicolumn{2}{c}{Component2}\\
		\cline{3-8}
		&  &$T$  &$N_{total}^{b}$ &$V_{\rm lsr}$ &FWHM &$V_{\rm lsr}$ &FWHM  \\
		& &(K)   &(cm$^{-2}$) &(km s$^{-1}$) &(km s$^{-1}$) &(km s$^{-1}$) &(km s$^{-1}$) 
	}
	\startdata
I14498-7	&1	&121.6(13.2) 	&1.6(1.0)E+15	&-50.9(0.1) 	&6.2(0.1) 	& & \\																							
I15520-18	&1	&104.6(15.5) 	&1.3(0.5)E+15	&-39.9(5.1) 	&7.2(0.1) 	& & \\																							
I15520-22	&1	&116.9(19.0) 	&9.5(1.1)E+14	&-42.9(0.1) 	&5.6(0.3) 	& & \\																							
I15596-19	&2	&97.6(15.9) 	&7.2(0.9)E+14	&-71.0(0.1) 	&7.5(0.6) 	&-75.0(0.1) 	&7.7(0.1) \\ 																
I16060-4	&1	&112.8(12.3) 	&6.0(0.6)E+15	&-94.4(0.1) 	&4.2(0.2) 	& & \\																					
I16060-7	&1	&100.7(11.2) 	&1.3(0.3)E+16	&-86.7(0.5) 	&5.2(1.0) 	& & \\																								
I16060-8	&1	&122.6(12.9) 	&2.1(0.5)E+16	&-94.6(0.8) 	&6.9(1.7) 	& & \\																								
I16060-9	&1	&112.2(12.4) 	&3.4(0.4)E+16	&-94.3(0.1) 	&4.9(0.3) 	& & \\
\enddata
	\tablecomments{\\
	$a$: The number of velocity components defined in the spectra.\\
	$b$: The column density are calculated by intergrating all different components.}
\end{deluxetable*}

\section{Spectra of detected species}\label{sec:B}
In this section, the spectra of all detected S-bearing species are presented. 
In each plot, the observed spectra were illustrated as a black solid line, and the LTE fitting of velocity components was denoted as colored curves. The red lines represent the summed fitting results of all velocity components. 
It is worth noting that the column densities of SO absorptions are estimated by equation (\ref{eq:1}). Therefore, the LTE fitting results presented in this section are only from the emissions of detected species. The absorptions have been excluded from the fittings.

\setcounter{figure}{0}
\renewcommand{\thefigure}{C\arabic{figure}}

\begin{figure}
	\plotone{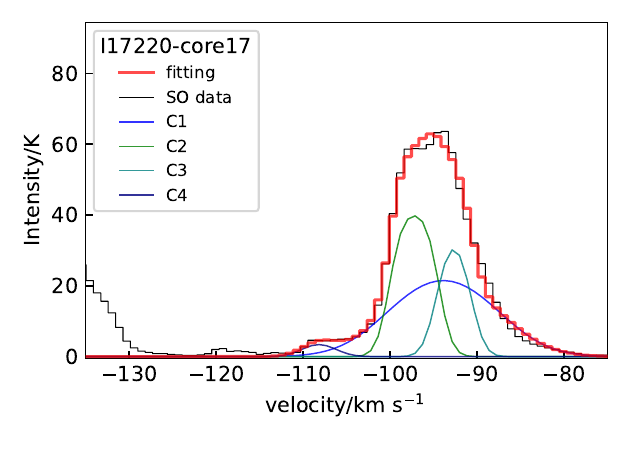}
	\caption{SO spectra and fitting results.The black stepped line represents observed data, the colored curves depict fitting results of decomposed velocity components, and the red stepped line shows the summed spectral of all fitting components. The complete figure set (159 images) can be found in the online journal.} 
	\label{fig:C1_SO}
\end{figure}

%
\begin{figure}
	\plotone{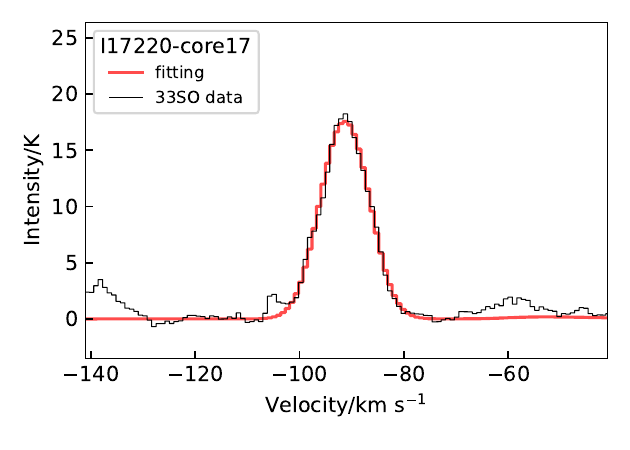}
	\caption{\ttso~spectra and fitting results. In the figure, the black stepped line represents observed data, the colored curves depict fitting results of decomposed velocity components, and the red stepped line shows the summed spectral of all fitting components. The complete figure set (36 images) can be found in the online journal.} 
	\label{fig:C2_33SO}
\end{figure}

%
\begin{figure}
	\plotone{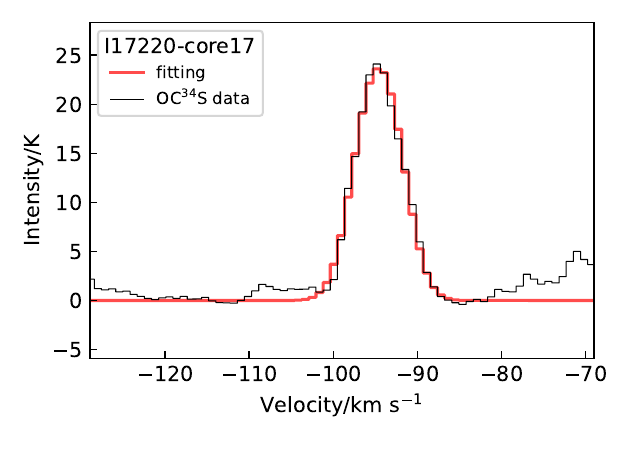}
	\caption{\octfs~spectra and fitting results. In the figure, the black stepped line represents observed data, the colored curves depict fitting results of decomposed velocity components, and the red stepped line shows the summed spectral of all fitting components. The complete figure set (39 images) can be found in the online journal.} 
	\label{fig:C3_OC34S}
\end{figure}

\begin{figure}
	\plotone{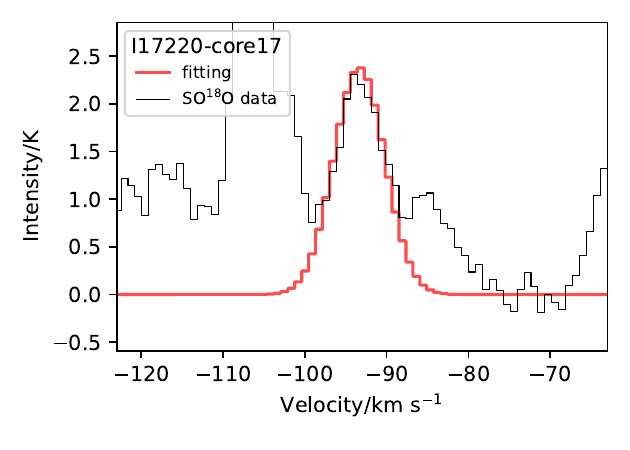}
	\caption{\seoo~spectra and fitting results. In the figure, the black stepped line represents observed data, the colored curves depict fitting results of decomposed velocity components, and the red stepped line shows the summed spectral of all fitting components. The complete figure set (13 images) can be found in the online journal.} 
	\label{fig:C5_SO18O}
\end{figure}

\begin{figure}
	\plotone{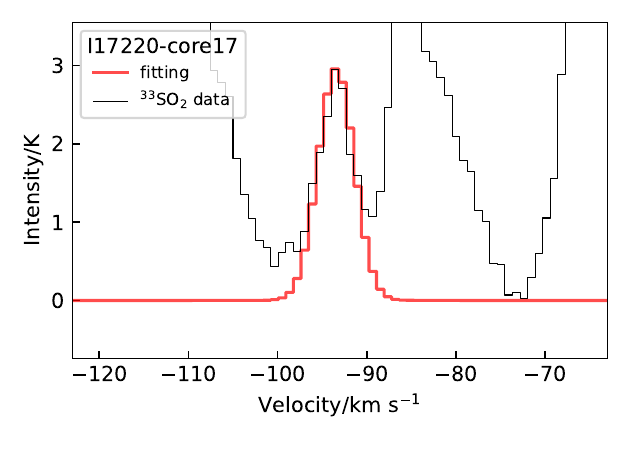}
	\caption{\ttsot~spectra and fitting results. The black stepped line represents observed data, the colored curves depict fitting results of decomposed velocity components, and the red stepped line shows the summed spectral of all fitting components. The complete figure set (15 images) can be found in the online journal.} 
	\label{fig:C6_33SO2}
\end{figure}

\begin{figure*}
	\plotone{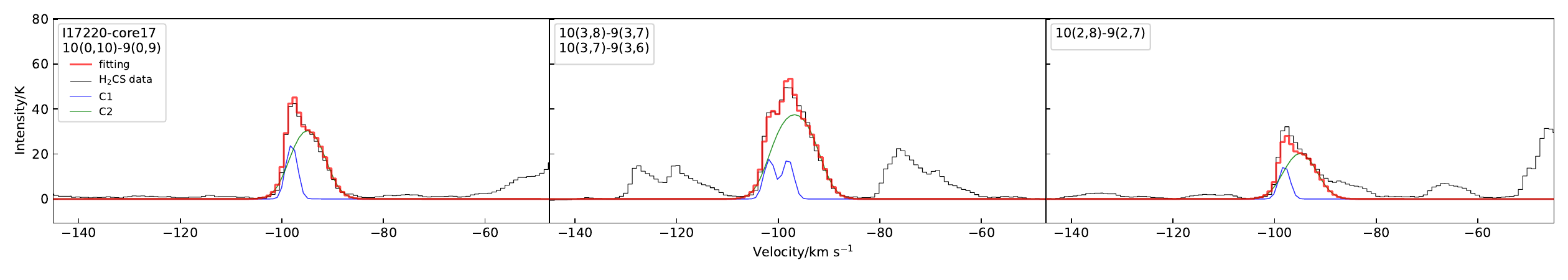}
	\caption{\htcs~spectra and fitting results.The black stepped line represents observed data, the colored curves depict fitting results of decomposed velocity components, and the red stepped line shows the summed spectral of all fitting components. The complete figure set (115 images) can be found in the online journal.} 
	\label{fig:C7_H2CS}
\end{figure*}

\begin{figure*}
	\plotone{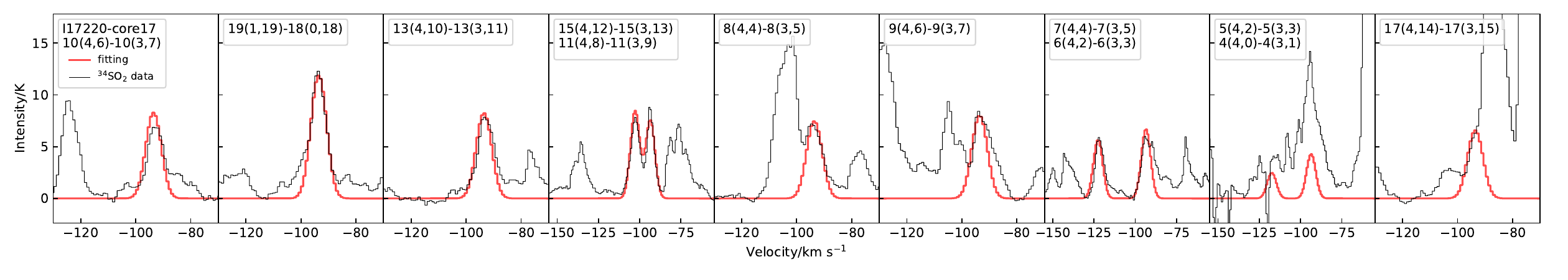}
	\caption{\tfsot~spectra and fitting results. The black stepped line represents observed data, the colored curves depict fitting results of decomposed velocity components, and the red stepped line shows the summed spectral of all fitting components. The complete figure set (27 images) can be found in the online journal.} 
	\label{fig:A6_34SO2}
\end{figure*}
\clearpage

\bibliography{reference}{}

\begin{thebibliography}{}
\expandafter\ifx\csname natexlab\endcsname\relax\def\natexlab#1{#1}\fi
\providecommand{\url}[1]{\href{#1}{#1}}
\providecommand{\dodoi}[1]{doi:~\href{http://doi.org/#1}{\nolinkurl{#1}}}
\providecommand{\doeprint}[1]{\href{http://ascl.net/#1}{\nolinkurl{http://ascl.net/#1}}}
\providecommand{\doarXiv}[1]{\href{https://arxiv.org/abs/#1}{\nolinkurl{https://arxiv.org/abs/#1}}}

\bibitem[{{Anderson} {et~al.}(2013){Anderson}, {Bergin}, {Maret}, \&
  {Wakelam}}]{Anderson2013ApJ...779..141A}
{Anderson}, D.~E., {Bergin}, E.~A., {Maret}, S., \& {Wakelam}, V. 2013, \apj,
  779, 141, \dodoi{10.1088/0004-637X/779/2/141}

\bibitem[{{Artur de la Villarmois} {et~al.}(2023){Artur de la Villarmois},
  {Guzm{\'a}n}, {Yang}, {Zhang}, \& {Sakai}}]{Artur2023A&A...678A.124A}
{Artur de la Villarmois}, E., {Guzm{\'a}n}, V.~V., {Yang}, Y.~L., {Zhang}, Y.,
  \& {Sakai}, N. 2023, \aap, 678, A124, \dodoi{10.1051/0004-6361/202346728}

\bibitem[{{Artur de la Villarmois} {et~al.}(2019){Artur de la Villarmois},
  {J{\o}rgensen}, {Kristensen}, {Bergin}, {Harsono}, {Sakai}, {van Dishoeck},
  \& {Yamamoto}}]{Artur2019A&A...626A..71A}
{Artur de la Villarmois}, E., {J{\o}rgensen}, J.~K., {Kristensen}, L.~E.,
  {et~al.} 2019, \aap, 626, A71, \dodoi{10.1051/0004-6361/201834877}

\bibitem[{{Astropy Collaboration} {et~al.}(2013){Astropy Collaboration},
  {Robitaille}, {Tollerud}, {Greenfield}, {Droettboom}, {Bray}, {Aldcroft},
  {Davis}, {Ginsburg}, {Price-Whelan}, {Kerzendorf}, {Conley}, {Crighton},
  {Barbary}, {Muna}, {Ferguson}, {Grollier}, {Parikh}, {Nair}, {Unther},
  {Deil}, {Woillez}, {Conseil}, {Kramer}, {Turner}, {Singer}, {Fox}, {Weaver},
  {Zabalza}, {Edwards}, {Azalee Bostroem}, {Burke}, {Casey}, {Crawford},
  {Dencheva}, {Ely}, {Jenness}, {Labrie}, {Lim}, {Pierfederici}, {Pontzen},
  {Ptak}, {Refsdal}, {Servillat}, \& {Streicher}}]{2013A&A...558A..33A}
{Astropy Collaboration}, {Robitaille}, T.~P., {Tollerud}, E.~J., {et~al.} 2013,
  \aap, 558, A33, \dodoi{10.1051/0004-6361/201322068}

\bibitem[{{Astropy Collaboration} {et~al.}(2018){Astropy Collaboration},
  {Price-Whelan}, {Sip{\H{o}}cz}, {G{\"u}nther}, {Lim}, {Crawford}, {Conseil},
  {Shupe}, {Craig}, {Dencheva}, {Ginsburg}, {VanderPlas}, {Bradley},
  {P{\'e}rez-Su{\'a}rez}, {de Val-Borro}, {Aldcroft}, {Cruz}, {Robitaille},
  {Tollerud}, {Ardelean}, {Babej}, {Bach}, {Bachetti}, {Bakanov}, {Bamford},
  {Barentsen}, {Barmby}, {Baumbach}, {Berry}, {Biscani}, {Boquien}, {Bostroem},
  {Bouma}, {Brammer}, {Bray}, {Breytenbach}, {Buddelmeijer}, {Burke},
  {Calderone}, {Cano Rodr{\'\i}guez}, {Cara}, {Cardoso}, {Cheedella}, {Copin},
  {Corrales}, {Crichton}, {D'Avella}, {Deil}, {Depagne}, {Dietrich}, {Donath},
  {Droettboom}, {Earl}, {Erben}, {Fabbro}, {Ferreira}, {Finethy}, {Fox},
  {Garrison}, {Gibbons}, {Goldstein}, {Gommers}, {Greco}, {Greenfield},
  {Groener}, {Grollier}, {Hagen}, {Hirst}, {Homeier}, {Horton}, {Hosseinzadeh},
  {Hu}, {Hunkeler}, {Ivezi{\'c}}, {Jain}, {Jenness}, {Kanarek}, {Kendrew},
  {Kern}, {Kerzendorf}, {Khvalko}, {King}, {Kirkby}, {Kulkarni}, {Kumar},
  {Lee}, {Lenz}, {Littlefair}, {Ma}, {Macleod}, {Mastropietro}, {McCully},
  {Montagnac}, {Morris}, {Mueller}, {Mumford}, {Muna}, {Murphy}, {Nelson},
  {Nguyen}, {Ninan}, {N{\"o}the}, {Ogaz}, {Oh}, {Parejko}, {Parley}, {Pascual},
  {Patil}, {Patil}, {Plunkett}, {Prochaska}, {Rastogi}, {Reddy Janga},
  {Sabater}, {Sakurikar}, {Seifert}, {Sherbert}, {Sherwood-Taylor}, {Shih},
  {Sick}, {Silbiger}, {Singanamalla}, {Singer}, {Sladen}, {Sooley},
  {Sornarajah}, {Streicher}, {Teuben}, {Thomas}, {Tremblay}, {Turner},
  {Terr{\'o}n}, {van Kerkwijk}, {de la Vega}, {Watkins}, {Weaver}, {Whitmore},
  {Woillez}, {Zabalza}, \& {Astropy Contributors}}]{2018AJ....156..123A}
{Astropy Collaboration}, {Price-Whelan}, A.~M., {Sip{\H{o}}cz}, B.~M., {et~al.}
  2018, \aj, 156, 123, \dodoi{10.3847/1538-3881/aabc4f}

\bibitem[{{Astropy Collaboration} {et~al.}(2022){Astropy Collaboration},
  {Price-Whelan}, {Lim}, {Earl}, {Starkman}, {Bradley}, {Shupe}, {Patil},
  {Corrales}, {Brasseur}, {N{\"o}the}, {Donath}, {Tollerud}, {Morris},
  {Ginsburg}, {Vaher}, {Weaver}, {Tocknell}, {Jamieson}, {van Kerkwijk},
  {Robitaille}, {Merry}, {Bachetti}, {G{\"u}nther}, {Aldcroft},
  {Alvarado-Montes}, {Archibald}, {B{\'o}di}, {Bapat}, {Barentsen},
  {Baz{\'a}n}, {Biswas}, {Boquien}, {Burke}, {Cara}, {Cara}, {Conroy},
  {Conseil}, {Craig}, {Cross}, {Cruz}, {D'Eugenio}, {Dencheva}, {Devillepoix},
  {Dietrich}, {Eigenbrot}, {Erben}, {Ferreira}, {Foreman-Mackey}, {Fox},
  {Freij}, {Garg}, {Geda}, {Glattly}, {Gondhalekar}, {Gordon}, {Grant},
  {Greenfield}, {Groener}, {Guest}, {Gurovich}, {Handberg}, {Hart},
  {Hatfield-Dodds}, {Homeier}, {Hosseinzadeh}, {Jenness}, {Jones}, {Joseph},
  {Kalmbach}, {Karamehmetoglu}, {Ka{\l}uszy{\'n}ski}, {Kelley}, {Kern},
  {Kerzendorf}, {Koch}, {Kulumani}, {Lee}, {Ly}, {Ma}, {MacBride}, {Maljaars},
  {Muna}, {Murphy}, {Norman}, {O'Steen}, {Oman}, {Pacifici}, {Pascual},
  {Pascual-Granado}, {Patil}, {Perren}, {Pickering}, {Rastogi}, {Roulston},
  {Ryan}, {Rykoff}, {Sabater}, {Sakurikar}, {Salgado}, {Sanghi}, {Saunders},
  {Savchenko}, {Schwardt}, {Seifert-Eckert}, {Shih}, {Jain}, {Shukla}, {Sick},
  {Simpson}, {Singanamalla}, {Singer}, {Singhal}, {Sinha}, {Sip{\H{o}}cz},
  {Spitler}, {Stansby}, {Streicher}, {{\v{S}}umak}, {Swinbank}, {Taranu},
  {Tewary}, {Tremblay}, {de Val-Borro}, {Van Kooten}, {Vasovi{\'c}}, {Verma},
  {de Miranda Cardoso}, {Williams}, {Wilson}, {Winkel}, {Wood-Vasey}, {Xue},
  {Yoachim}, {Zhang}, {Zonca}, \& {Astropy Project
  Contributors}}]{2022ApJ...935..167A}
{Astropy Collaboration}, {Price-Whelan}, A.~M., {Lim}, P.~L., {et~al.} 2022,
  \apj, 935, 167, \dodoi{10.3847/1538-4357/ac7c74}

\bibitem[{{Bachiller} \& {P{\'e}rez
  Guti{\'e}rrez}(1997)}]{Bachiller1997ApJ...487L..93B}
{Bachiller}, R., \& {P{\'e}rez Guti{\'e}rrez}, M. 1997, \apjl, 487, L93,
  \dodoi{10.1086/310877}

\bibitem[{{Bachiller} {et~al.}(2001){Bachiller}, {P{\'e}rez Guti{\'e}rrez},
  {Kumar}, \& {Tafalla}}]{Bachiller2001A&A...372..899B}
{Bachiller}, R., {P{\'e}rez Guti{\'e}rrez}, M., {Kumar}, M.~S.~N., \&
  {Tafalla}, M. 2001, \aap, 372, 899, \dodoi{10.1051/0004-6361:20010519}

\bibitem[{{Baug} {et~al.}(2021){Baug}, {Wang}, {Liu}, {Wu}, {Li}, {Zhang},
  {Tang}, {Goldsmith}, {Liu}, {Tej}, {Bronfman}, {Kim}, {Li}, {Lee},
  {Tatematsu}, {Hirota}, \& {Toth}}]{Baug2021MNRAS.507.4316B}
{Baug}, T., {Wang}, K., {Liu}, T., {et~al.} 2021, \mnras, 507, 4316,
  \dodoi{10.1093/mnras/stab1902}

\bibitem[{{Boogert} {et~al.}(1997){Boogert}, {Schutte}, {Helmich}, {Tielens},
  \& {Wooden}}]{Boogert1997A&A...317..929B}
{Boogert}, A.~C.~A., {Schutte}, W.~A., {Helmich}, F.~P., {Tielens},
  A.~G.~G.~M., \& {Wooden}, D.~H. 1997, \aap, 317, 929

\bibitem[{{Booth} {et~al.}(2023){Booth}, {Ilee}, {Walsh}, {Kama}, {Keyte}, {van
  Dishoeck}, \& {Nomura}}]{Booth2023A&A...669A..53B}
{Booth}, A.~S., {Ilee}, J.~D., {Walsh}, C., {et~al.} 2023, \aap, 669, A53,
  \dodoi{10.1051/0004-6361/202244472}

\bibitem[{{Bouscasse} {et~al.}(2022){Bouscasse}, {Csengeri}, {Belloche},
  {Wyrowski}, {Bontemps}, {G{\"u}sten}, \&
  {Menten}}]{Bouscasse2022A&A...662A..32B}
{Bouscasse}, L., {Csengeri}, T., {Belloche}, A., {et~al.} 2022, \aap, 662, A32,
  \dodoi{10.1051/0004-6361/202140519}

\bibitem[{{Braatz}(2020)}]{almaguide}
{Braatz}, J. 2020, ALMA Cycle 8 Proposer's Guide, ALMA Doc. 8.2 v1.0.
\newblock
  \url{https://almascience.nrao.edu/documents-and-tools/cycle8/alma-proposers-guide}

\bibitem[{{Brogan} {et~al.}(2018){Brogan}, {Hunter}, \&
  {Fomalont}}]{Brogan2018arXiv180505266B}
{Brogan}, C.~L., {Hunter}, T.~R., \& {Fomalont}, E.~B. 2018, arXiv e-prints,
  arXiv:1805.05266, \dodoi{10.48550/arXiv.1805.05266}

\bibitem[{{CASA Team} {et~al.}(2022){CASA Team}, {Bean}, {Bhatnagar}, {Castro},
  {Donovan Meyer}, {Emonts}, {Garcia}, {Garwood}, {Golap}, {Gonzalez Villalba},
  {Harris}, {Hayashi}, {Hoskins}, {Hsieh}, {Jagannathan}, {Kawasaki},
  {Keimpema}, {Kettenis}, {Lopez}, {Marvil}, {Masters}, {McNichols},
  {Mehringer}, {Miel}, {Moellenbrock}, {Montesino}, {Nakazato}, {Ott}, {Petry},
  {Pokorny}, {Raba}, {Rau}, {Schiebel}, {Schweighart}, {Sekhar}, {Shimada},
  {Small}, {Steeb}, {Sugimoto}, {Suoranta}, {Tsutsumi}, {van Bemmel},
  {Verkouter}, {Wells}, {Xiong}, {Szomoru}, {Griffith}, {Glendenning}, \&
  {Kern}}]{2022PASP..134k4501C}
{CASA Team}, {Bean}, B., {Bhatnagar}, S., {et~al.} 2022, \pasp, 134, 114501,
  \dodoi{10.1088/1538-3873/ac9642}

\bibitem[{{Charnley}(1997)}]{Charnley1997ApJ...481..396C}
{Charnley}, S.~B. 1997, \apj, 481, 396, \dodoi{10.1086/304011}

\bibitem[{{Chen} {et~al.}(2024){Chen}, {Qin}, {Liu}, {Liu}, {Liu}, {Liu},
  {Shi}, {Li}, {Tang}, {Zhang}, {Tatematsu}, {Li}, {Xu}, {Wu}, \&
  {Yang}}]{Chen2024ApJ...962...13C}
{Chen}, L., {Qin}, S.-L., {Liu}, T., {et~al.} 2024, \apj, 962, 13,
  \dodoi{10.3847/1538-4357/ad110f}

\bibitem[{{Coletta} {et~al.}(2020){Coletta}, {Fontani}, {Rivilla}, {Mininni},
  {Colzi}, {S{\'a}nchez-Monge}, \& {Beltr{\'a}n}}]{2020A&A...641A..54C}
{Coletta}, A., {Fontani}, F., {Rivilla}, V.~M., {et~al.} 2020, \aap, 641, A54,
  \dodoi{10.1051/0004-6361/202038212}

\bibitem[{{Druard} \& {Wakelam}(2012)}]{Druard2012MNRAS.426..354D}
{Druard}, C., \& {Wakelam}, V. 2012, \mnras, 426, 354,
  \dodoi{10.1111/j.1365-2966.2012.21712.x}

\bibitem[{{Fa{\'u}ndez} {et~al.}(2004){Fa{\'u}ndez}, {Bronfman}, {Garay},
  {Chini}, {Nyman}, \& {May}}]{Faundez2004A&A...426...97F}
{Fa{\'u}ndez}, S., {Bronfman}, L., {Garay}, G., {et~al.} 2004, \aap, 426, 97,
  \dodoi{10.1051/0004-6361:20035755}

\bibitem[{{Fontani} {et~al.}(2023){Fontani}, {Roueff}, {Colzi}, \&
  {Caselli}}]{Fontani2023A&A...680A..58F}
{Fontani}, F., {Roueff}, E., {Colzi}, L., \& {Caselli}, P. 2023, \aap, 680,
  A58, \dodoi{10.1051/0004-6361/202347565}

\bibitem[{{Frau} {et~al.}(2010){Frau}, {Girart}, {Beltr{\'a}n}, {Morata},
  {Masqu{\'e}}, {Busquet}, {Alves}, {S{\'a}nchez-Monge}, {Estalella}, \&
  {Franco}}]{Frau2010ApJ...723.1665F}
{Frau}, P., {Girart}, J.~M., {Beltr{\'a}n}, M.~T., {et~al.} 2010, \apj, 723,
  1665, \dodoi{10.1088/0004-637X/723/2/1665}

\bibitem[{{Fuente} {et~al.}(2021){Fuente}, {Trevi{\~n}o-Morales},
  {Alonso-Albi}, {S{\'a}nchez-Monge}, {Rivi{\`e}re-Marichalar}, \&
  {Navarro-Almaida}}]{Fuente2021MNRAS.507.1886F}
{Fuente}, A., {Trevi{\~n}o-Morales}, S.~P., {Alonso-Albi}, T., {et~al.} 2021,
  \mnras, 507, 1886, \dodoi{10.1093/mnras/stab2216}

\bibitem[{{Fuente} {et~al.}(2016){Fuente}, {Cernicharo}, {Roueff}, {Gerin},
  {Pety}, {Marcelino}, {Bachiller}, {Lefloch}, {Roncero}, \&
  {Aguado}}]{Fuente2016A&A...593A..94F}
{Fuente}, A., {Cernicharo}, J., {Roueff}, E., {et~al.} 2016, \aap, 593, A94,
  \dodoi{10.1051/0004-6361/201628285}

\bibitem[{{Garc{\'\i}a-Rojas} {et~al.}(2006){Garc{\'\i}a-Rojas}, {Esteban},
  {Peimbert}, {Costado}, {Rodr{\'\i}guez}, {Peimbert}, \&
  {Ruiz}}]{2006MNRAS.368..253G}
{Garc{\'\i}a-Rojas}, J., {Esteban}, C., {Peimbert}, M., {et~al.} 2006, \mnras,
  368, 253, \dodoi{10.1111/j.1365-2966.2006.10105.x}

\bibitem[{{Geballe} {et~al.}(1985){Geballe}, {Baas}, {Greenberg}, \&
  {Schutte}}]{Geballe1985A&A...146L...6G}
{Geballe}, T.~R., {Baas}, F., {Greenberg}, J.~M., \& {Schutte}, W. 1985, \aap,
  146, L6

\bibitem[{{Gieser} {et~al.}(2021){Gieser}, {Beuther}, {Semenov}, {Ahmadi},
  {Suri}, {M{\"o}ller}, {Beltr{\'a}n}, {Klaassen}, {Zhang}, {Urquhart},
  {Henning}, {Feng}, {Galv{\'a}n-Madrid}, {de Souza Magalh{\~a}es},
  {Moscadelli}, {Longmore}, {Leurini}, {Kuiper}, {Peters}, {Menten},
  {Csengeri}, {Fuller}, {Wyrowski}, {Lumsden}, {S{\'a}nchez-Monge}, {Maud},
  {Linz}, {Palau}, {Schilke}, {Pety}, {Pudritz}, {Winters}, \&
  {Pi{\'e}tu}}]{Gieser2021A&A...648A..66G}
{Gieser}, C., {Beuther}, H., {Semenov}, D., {et~al.} 2021, \aap, 648, A66,
  \dodoi{10.1051/0004-6361/202039670}

\bibitem[{{Goldsmith}(2001)}]{Goldsmith2001ApJ...557..736G}
{Goldsmith}, P.~F. 2001, \apj, 557, 736, \dodoi{10.1086/322255}

\bibitem[{{Hatchell} {et~al.}(1998){Hatchell}, {Thompson}, {Millar}, \&
  {MacDonald}}]{Hatchell1998A&A...338..713H}
{Hatchell}, J., {Thompson}, M.~A., {Millar}, T.~J., \& {MacDonald}, G.~H. 1998,
  \aap, 338, 713

\bibitem[{{Holdship} {et~al.}(2019){Holdship}, {Jimenez-Serra}, {Viti},
  {Codella}, {Benedettini}, {Fontani}, {Tafalla}, {Bachiller}, {Ceccarelli}, \&
  {Podio}}]{Holdship2019ApJ...878...64H}
{Holdship}, J., {Jimenez-Serra}, I., {Viti}, S., {et~al.} 2019, \apj, 878, 64,
  \dodoi{10.3847/1538-4357/ab1cb5}

\bibitem[{{Hony} {et~al.}(2002){Hony}, {Bouwman}, {Keller}, \&
  {Waters}}]{Hony2002A&A...393L.103H}
{Hony}, S., {Bouwman}, J., {Keller}, L.~P., \& {Waters}, L.~B.~F.~M. 2002,
  \aap, 393, L103, \dodoi{10.1051/0004-6361:20021260}

\bibitem[{{Howk} {et~al.}(2006){Howk}, {Sembach}, \&
  {Savage}}]{How2006ApJ...637..333H}
{Howk}, J.~C., {Sembach}, K.~R., \& {Savage}, B.~D. 2006, \apj, 637, 333,
  \dodoi{10.1086/497352}

\bibitem[{{Jim{\'e}nez-Escobar} \& {Mu{\~n}oz
  Caro}(2011)}]{Jim2011A&A...536A..91J}
{Jim{\'e}nez-Escobar}, A., \& {Mu{\~n}oz Caro}, G.~M. 2011, \aap, 536, A91,
  \dodoi{10.1051/0004-6361/201014821}

\bibitem[{{Jim{\'e}nez-Serra} {et~al.}(2021){Jim{\'e}nez-Serra}, {Vasyunin},
  {Spezzano}, {Caselli}, {Cosentino}, \& {Viti}}]{2021ApJ...917...44J}
{Jim{\'e}nez-Serra}, I., {Vasyunin}, A.~I., {Spezzano}, S., {et~al.} 2021,
  \apj, 917, 44, \dodoi{10.3847/1538-4357/ac024c}

\bibitem[{{Kama} {et~al.}(2019){Kama}, {Shorttle}, {Jermyn}, {Folsom},
  {Furuya}, {Bergin}, {Walsh}, \& {Keller}}]{Kama2019ApJ...885..114K}
{Kama}, M., {Shorttle}, O., {Jermyn}, A.~S., {et~al.} 2019, \apj, 885, 114,
  \dodoi{10.3847/1538-4357/ab45f8}

\bibitem[{{Keller} {et~al.}(2002){Keller}, {Hony}, {Bradley}, {Molster},
  {Waters}, {Bouwman}, {de Koter}, {Brownlee}, {Flynn}, {Henning}, \&
  {Mutschke}}]{Keller2002Natur.417..148K}
{Keller}, L.~P., {Hony}, S., {Bradley}, J.~P., {et~al.} 2002, \nat, 417, 148,
  \dodoi{10.1038/417148a}

\bibitem[{{Laas} \& {Caselli}(2019)}]{Laas2019A&A...624A.108L}
{Laas}, J.~C., \& {Caselli}, P. 2019, \aap, 624, A108,
  \dodoi{10.1051/0004-6361/201834446}

\bibitem[{{Larimer}(1967)}]{Larimer1967GeCoA..31.1215L}
{Larimer}, J.~W. 1967, \gca, 31, 1215, \dodoi{10.1016/S0016-7037(67)80013-9}

\bibitem[{{Lee} {et~al.}(2018){Lee}, {Li}, {Hirano}, {Shang}, {Ho}, \&
  {Zhang}}]{Lee2018ApJ...863...94L}
{Lee}, C.-F., {Li}, Z.-Y., {Hirano}, N., {et~al.} 2018, \apj, 863, 94,
  \dodoi{10.3847/1538-4357/aad2da}

\bibitem[{{Liu} {et~al.}(2023){Liu}, {Qin}, {Liu}, {Tang}, {Liu}, {Chen}, {Li},
  {Shi}, {Li}, {Zhang}, {Tatematsu}, {Xu}, \& {Wu}}]{Liu2023ApJ...958..174L}
{Liu}, M., {Qin}, S.-L., {Liu}, T., {et~al.} 2023, \apj, 958, 174,
  \dodoi{10.3847/1538-4357/ad00aa}

\bibitem[{{Liu} {et~al.}(2011){Liu}, {Wu}, {Liu}, {Qin}, {Su}, {Chen}, \&
  {Ren}}]{Liu2011ApJ...730..102L}
{Liu}, T., {Wu}, Y., {Liu}, S.-Y., {et~al.} 2011, \apj, 730, 102,
  \dodoi{10.1088/0004-637X/730/2/102}

\bibitem[{{Liu} {et~al.}(2016){Liu}, {Kim}, {Yoo}, {Liu}, {Tatematsu}, {Qin},
  {Zhang}, {Wu}, {Wang}, {Goldsmith}, {Juvela}, {Lee}, {T{\'o}th}, {Mardones},
  {Garay}, {Bronfman}, {Cunningham}, {Li}, {Lo}, {Ristorcelli}, \&
  {Schnee}}]{Liu2016ApJ...829...59L}
{Liu}, T., {Kim}, K.-T., {Yoo}, H., {et~al.} 2016, \apj, 829, 59,
  \dodoi{10.3847/0004-637X/829/2/59}

\bibitem[{{Lodders}(2003)}]{Lodders2003ApJ...591.1220L}
{Lodders}, K. 2003, \apj, 591, 1220, \dodoi{10.1086/375492}

\bibitem[{{Luo} {et~al.}(2019){Luo}, {Feng}, {Li}, {Qin}, {Peng}, {Tang},
  {Ren}, \& {Shi}}]{Luo2019ApJ...885...82L}
{Luo}, G., {Feng}, S., {Li}, D., {et~al.} 2019, \apj, 885, 82,
  \dodoi{10.3847/1538-4357/ab45ef}

\bibitem[{{Mangum} \& {Shirley}(2015)}]{Mangum2015}
{Mangum}, J.~G., \& {Shirley}, Y.~L. 2015, \pasp, 127, 266,
  \dodoi{10.1086/680323}

\bibitem[{{Mart{\'\i}n} {et~al.}(2019){Mart{\'\i}n}, {Mart{\'\i}n-Pintado},
  {Blanco-S{\'a}nchez}, {Rivilla}, {Rodr{\'\i}guez-Franco}, \&
  {Rico-Villas}}]{2019A&A...631A.159M}
{Mart{\'\i}n}, S., {Mart{\'\i}n-Pintado}, J., {Blanco-S{\'a}nchez}, C.,
  {et~al.} 2019, \aap, 631, A159, \dodoi{10.1051/0004-6361/201936144}

\bibitem[{{Mart{\'\i}n-Hern{\'a}ndez}
  {et~al.}(2002){Mart{\'\i}n-Hern{\'a}ndez}, {Peeters}, {Morisset}, {Tielens},
  {Cox}, {Roelfsema}, {Baluteau}, {Schaerer}, {Mathis}, {Damour}, {Churchwell},
  \& {Kessler}}]{2002A&A...381..606M}
{Mart{\'\i}n-Hern{\'a}ndez}, N.~L., {Peeters}, E., {Morisset}, C., {et~al.}
  2002, \aap, 381, 606, \dodoi{10.1051/0004-6361:20011504}

\bibitem[{{McClure} {et~al.}(2023){McClure}, {Rocha}, {Pontoppidan}, {Crouzet},
  {Chu}, {Dartois}, {Lamberts}, {Noble}, {Pendleton}, {Perotti}, {Qasim},
  {Rachid}, {Smith}, {Sun}, {Beck}, {Boogert}, {Brown}, {Caselli}, {Charnley},
  {Cuppen}, {Dickinson}, {Drozdovskaya}, {Egami}, {Erkal}, {Fraser}, {Garrod},
  {Harsono}, {Ioppolo}, {Jim{\'e}nez-Serra}, {Jin}, {J{\o}rgensen},
  {Kristensen}, {Lis}, {McCoustra}, {McGuire}, {Melnick}, {{\~A}-berg},
  {Palumbo}, {Shimonishi}, {Sturm}, {van Dishoeck}, \&
  {Linnartz}}]{McClure2023NatAs...7..431M}
{McClure}, M.~K., {Rocha}, W.~R.~M., {Pontoppidan}, K.~M., {et~al.} 2023,
  Nature Astronomy, 7, 431, \dodoi{10.1038/s41550-022-01875-w}

\bibitem[{{McDowell}(1988)}]{1988JChPh..88..356M}
{McDowell}, R.~S. 1988, \jcp, 88, 356, \dodoi{10.1063/1.454608}

\bibitem[{{Minh} {et~al.}(2018){Minh}, {Liu}, {Galva{\'n}-Madrid}, {Sahu},
  {He}, \& {Hasegawa}}]{Minh2018ApJ...864..102M}
{Minh}, Y.~C., {Liu}, H.~B., {Galva{\'n}-Madrid}, R., {et~al.} 2018, \apj, 864,
  102, \dodoi{10.3847/1538-4357/aad909}

\bibitem[{{Minh} {et~al.}(2011){Minh}, {Liu}, {Chen}, \&
  {Su}}]{Minh2011ApJ...737L..25M}
{Minh}, Y.~C., {Liu}, S.~Y., {Chen}, H.~R., \& {Su}, Y.~N. 2011, \apjl, 737,
  L25, \dodoi{10.1088/2041-8205/737/1/L25}

\bibitem[{{Minh} {et~al.}(1990){Minh}, {Ziurys}, {Irvine}, \&
  {McGonagle}}]{Minh1990ApJ...360..136M}
{Minh}, Y.~C., {Ziurys}, L.~M., {Irvine}, W.~M., \& {McGonagle}, D. 1990, \apj,
  360, 136, \dodoi{10.1086/169103}

\bibitem[{{Mininni} {et~al.}(2023){Mininni}, {Beltr{\'a}n}, {Colzi}, {Rivilla},
  {Fontani}, {Lorenzani}, {L{\'o}pez-Gallifa}, {Viti}, {S{\'a}nchez-Monge},
  {Schilke}, \& {Testi}}]{2023A&A...677A..15M}
{Mininni}, C., {Beltr{\'a}n}, M.~T., {Colzi}, L., {et~al.} 2023, \aap, 677,
  A15, \dodoi{10.1051/0004-6361/202245277}

\bibitem[{{M{\"o}ller} {et~al.}(2021){M{\"o}ller}, {Schilke}, {Schmiedeke},
  {Bergin}, {Lis}, {S{\'a}nchez-Monge}, {Schw{\"o}rer}, \&
  {Comito}}]{Moller2021A&A...651A...9M}
{M{\"o}ller}, T., {Schilke}, P., {Schmiedeke}, A., {et~al.} 2021, \aap, 651,
  A9, \dodoi{10.1051/0004-6361/202040203}

\bibitem[{{M{\"u}ller} {et~al.}(2005){M{\"u}ller}, {Schl{\"o}der}, {Stutzki},
  \& {Winnewisser}}]{CDMS}
{M{\"u}ller}, H. S.~P., {Schl{\"o}der}, F., {Stutzki}, J., \& {Winnewisser}, G.
  2005, Journal of Molecular Structure, 742, 215,
  \dodoi{10.1016/j.molstruc.2005.01.027}

\bibitem[{{Navarro-Almaida} {et~al.}(2020){Navarro-Almaida}, {Le Gal},
  {Fuente}, {Rivi{\`e}re-Marichalar}, {Wakelam}, {Cazaux}, {Caselli}, {Laas},
  {Alonso-Albi}, {Loison}, {Gerin}, {Kramer}, {Roueff}, {Bachiller},
  {Commer{\c{c}}on}, {Friesen}, {Garc{\'\i}a-Burillo}, {Goicoechea},
  {Giuliano}, {Jim{\'e}nez-Serra}, {Kirk}, {Lattanzi}, {Malinen}, {Marcelino},
  {Mart{\'\i}n-Dom{\`e}nech}, {Mu{\~n}oz Caro}, {Pineda}, {Tercero},
  {Trevi{\~n}o-Morales}, {Roncero}, {Hacar}, {Tafalla}, \&
  {Ward-Thompson}}]{Navarro2020A&A...637A..39N}
{Navarro-Almaida}, D., {Le Gal}, R., {Fuente}, A., {et~al.} 2020, \aap, 637,
  A39, \dodoi{10.1051/0004-6361/201937180}

\bibitem[{{Oppenheimer} \& {Dalgarno}(1974)}]{Oppenheimer1974ApJ...187..231O}
{Oppenheimer}, M., \& {Dalgarno}, A. 1974, \apj, 187, 231,
  \dodoi{10.1086/152618}

\bibitem[{{Oya} {et~al.}(2019){Oya}, {L{\'o}pez-Sepulcre}, {Sakai}, {Watanabe},
  {Higuchi}, {Hirota}, {Aikawa}, {Sakai}, {Ceccarelli}, {Lefloch}, {Caux},
  {Vastel}, {Kahane}, \& {Yamamoto}}]{Oya2019ApJ...881..112O}
{Oya}, Y., {L{\'o}pez-Sepulcre}, A., {Sakai}, N., {et~al.} 2019, \apj, 881,
  112, \dodoi{10.3847/1538-4357/ab2b97}

\bibitem[{{Palumbo} {et~al.}(1995){Palumbo}, {Tielens}, \&
  {Tokunaga}}]{Palumbo1995ApJ...449..674P}
{Palumbo}, M.~E., {Tielens}, A.~G.~G.~M., \& {Tokunaga}, A.~T. 1995, \apj, 449,
  674, \dodoi{10.1086/176088}

\bibitem[{{Penzias} {et~al.}(1971){Penzias}, {Solomon}, {Wilson}, \&
  {Jefferts}}]{Penzias1971ApJ...168L..53P}
{Penzias}, A.~A., {Solomon}, P.~M., {Wilson}, R.~W., \& {Jefferts}, K.~B. 1971,
  \apjl, 168, L53, \dodoi{10.1086/180784}

\bibitem[{{Pineau des Forets} {et~al.}(1993){Pineau des Forets}, {Roueff},
  {Schilke}, \& {Flower}}]{Pineau1993MNRAS.262..915P}
{Pineau des Forets}, G., {Roueff}, E., {Schilke}, P., \& {Flower}, D.~R. 1993,
  \mnras, 262, 915, \dodoi{10.1093/mnras/262.4.915}

\bibitem[{{Qin} {et~al.}(2022){Qin}, {Liu}, {Liu}, {Goldsmith}, {Li}, {Zhang},
  {Liu}, {Wu}, {Bronfman}, {Juvela}, {Lee}, {Garay}, {Zhang}, {He}, {Hsu},
  {Shen}, {Lee}, {Wang}, {Tang}, {Tang}, {Zhang}, {Yue}, {Xue}, {Li}, {Peng},
  {Dutta}, {Ge}, {Xu}, {Chen}, {Baug}, {Dewangan}, \&
  {Tej}}]{Qin2022MNRAS.511.3463Q}
{Qin}, S.-L., {Liu}, T., {Liu}, X., {et~al.} 2022, \mnras, 511, 3463,
  \dodoi{10.1093/mnras/stac219}

\bibitem[{{Richards} {et~al.}(2022){Richards}, {Moravec}, {Etoka}, {Fomalont},
  {P{\'e}rez-S{\'a}nchez}, {Toribio}, \& {Laing}}]{Richards2022arXiv220705591R}
{Richards}, A.~M.~S., {Moravec}, E., {Etoka}, S., {et~al.} 2022, arXiv
  e-prints, arXiv:2207.05591, \dodoi{10.48550/arXiv.2207.05591}

\bibitem[{{Rivi{\`e}re-Marichalar} {et~al.}(2019){Rivi{\`e}re-Marichalar},
  {Fuente}, {Goicoechea}, {Pety}, {Le Gal}, {Gratier}, {Guzm{\'a}n}, {Roueff},
  {Loison}, {Wakelam}, \& {Gerin}}]{Rivi2019A&A...628A..16R}
{Rivi{\`e}re-Marichalar}, P., {Fuente}, A., {Goicoechea}, J.~R., {et~al.} 2019,
  \aap, 628, A16, \dodoi{10.1051/0004-6361/201935354}

\bibitem[{{Rivilla} {et~al.}(2020){Rivilla}, {Colzi}, {Fontani}, {Melosso},
  {Caselli}, {Bizzocchi}, {Tamassia}, \& {Dore}}]{2020MNRAS.496.1990R}
{Rivilla}, V.~M., {Colzi}, L., {Fontani}, F., {et~al.} 2020, \mnras, 496, 1990,
  \dodoi{10.1093/mnras/staa1616}

\bibitem[{{Rocha} {et~al.}(2024){Rocha}, {van Dishoeck}, {Ressler}, {van
  Gelder}, {Slavicinska}, {Brunken}, {Linnartz}, {Ray}, {Beuther}, {Caratti o
  Garatti}, {Geers}, {Kavanagh}, {Klaassen}, {Justtanont}, {Chen}, {Francis},
  {Gieser}, {Perotti}, {Tychoniec}, {Barsony}, {Majumdar}, {le Gouellec},
  {Chu}, {Lew}, {Henning}, \& {Wright}}]{Rocha2024A&A...683A.124R}
{Rocha}, W.~R.~M., {van Dishoeck}, E.~F., {Ressler}, M.~E., {et~al.} 2024,
  \aap, 683, A124, \dodoi{10.1051/0004-6361/202348427}

\bibitem[{{Ruffle} {et~al.}(1999){Ruffle}, {Hartquist}, {Caselli}, \&
  {Williams}}]{Ruffle1999MNRAS.306..691R}
{Ruffle}, D.~P., {Hartquist}, T.~W., {Caselli}, P., \& {Williams}, D.~A. 1999,
  \mnras, 306, 691, \dodoi{10.1046/j.1365-8711.1999.02562.x}

\bibitem[{{Sakai} {et~al.}(2014){Sakai}, {Sakai}, {Hirota}, {Watanabe},
  {Ceccarelli}, {Kahane}, {Bottinelli}, {Caux}, {Demyk}, {Vastel}, {Coutens},
  {Taquet}, {Ohashi}, {Takakuwa}, {Yen}, {Aikawa}, \&
  {Yamamoto}}]{Sakai2014Natur.507...78S}
{Sakai}, N., {Sakai}, T., {Hirota}, T., {et~al.} 2014, \nat, 507, 78,
  \dodoi{10.1038/nature13000}

\bibitem[{{Sakai} {et~al.}(2017){Sakai}, {Oya}, {Higuchi}, {Aikawa}, {Hanawa},
  {Ceccarelli}, {Lefloch}, {L{\'o}pez-Sepulcre}, {Watanabe}, {Sakai}, {Hirota},
  {Caux}, {Vastel}, {Kahane}, \& {Yamamoto}}]{Sakai2017MNRAS.467L..76S}
{Sakai}, N., {Oya}, Y., {Higuchi}, A.~E., {et~al.} 2017, \mnras, 467, L76,
  \dodoi{10.1093/mnrasl/slx002}

\bibitem[{{Savage} \& {Sembach}(1996)}]{Savage1996ApJ...470..893S}
{Savage}, B.~D., \& {Sembach}, K.~R. 1996, \apj, 470, 893,
  \dodoi{10.1086/177919}

\bibitem[{{Scappini} {et~al.}(2003){Scappini}, {Cecchi-Pestellini}, {Smith},
  {Klemperer}, \& {Dalgarno}}]{Scappini2003MNRAS.341..657S}
{Scappini}, F., {Cecchi-Pestellini}, C., {Smith}, H., {Klemperer}, W., \&
  {Dalgarno}, A. 2003, \mnras, 341, 657,
  \dodoi{10.1046/j.1365-8711.2003.06443.x}

\bibitem[{{Semenov} {et~al.}(2018){Semenov}, {Favre}, {Fedele}, {Guilloteau},
  {Teague}, {Henning}, {Dutrey}, {Chapillon}, {Hersant}, \&
  {Pi{\'e}tu}}]{Semenov2018A&A...617A..28S}
{Semenov}, D., {Favre}, C., {Fedele}, D., {et~al.} 2018, \aap, 617, A28,
  \dodoi{10.1051/0004-6361/201832980}

\bibitem[{{Shimajiri} {et~al.}(2015){Shimajiri}, {Sakai}, {Kitamura},
  {Tsukagoshi}, {Saito}, {Nakamura}, {Momose}, {Takakuwa}, {Yamaguchi},
  {Sakai}, {Yamamoto}, \& {Kawabe}}]{Shimajiri2015ApJS..221...31S}
{Shimajiri}, Y., {Sakai}, T., {Kitamura}, Y., {et~al.} 2015, \apjs, 221, 31,
  \dodoi{10.1088/0067-0049/221/2/31}

\bibitem[{{Tabone} {et~al.}(2017){Tabone}, {Cabrit}, {Bianchi}, {Ferreira},
  {Pineau des For{\^e}ts}, {Codella}, {Gusdorf}, {Gueth}, {Podio}, \&
  {Chapillon}}]{Tabone2017A&A...607L...6T}
{Tabone}, B., {Cabrit}, S., {Bianchi}, E., {et~al.} 2017, \aap, 607, L6,
  \dodoi{10.1051/0004-6361/201731691}

\bibitem[{{Taquet} {et~al.}(2020){Taquet}, {Codella}, {De Simone},
  {L{\'o}pez-Sepulcre}, {Pineda}, {Segura-Cox}, {Ceccarelli}, {Caselli},
  {Gusdorf}, {Persson}, {Alves}, {Caux}, {Favre}, {Fontani}, {Neri}, {Oya},
  {Sakai}, {Vastel}, {Yamamoto}, {Bachiller}, {Balucani}, {Bianchi},
  {Bizzocchi}, {Chac{\'o}n-Tanarro}, {Dulieu}, {Enrique-Romero}, {Feng},
  {Holdship}, {Lefloch}, {Jaber Al-Edhari}, {Jim{\'e}nez-Serra}, {Kahane},
  {Lattanzi}, {Ospina-Zamudio}, {Podio}, {Punanova}, {Rimola}, {Sims},
  {Spezzano}, {Testi}, {Theul{\'e}}, {Ugliengo}, {Vasyunin}, {Vazart}, {Viti},
  \& {Witzel}}]{Taquet2020A&A...637A..63T}
{Taquet}, V., {Codella}, C., {De Simone}, M., {et~al.} 2020, \aap, 637, A63,
  \dodoi{10.1051/0004-6361/201937072}

\bibitem[{{Tieftrunk} {et~al.}(1994){Tieftrunk}, {Pineau des Forets},
  {Schilke}, \& {Walmsley}}]{Tieftrunk1994A&A...289..579T}
{Tieftrunk}, A., {Pineau des Forets}, G., {Schilke}, P., \& {Walmsley}, C.~M.
  1994, \aap, 289, 579

\bibitem[{{Tychoniec} {et~al.}(2021){Tychoniec}, {van Dishoeck}, {van't Hoff},
  {van Gelder}, {Tabone}, {Chen}, {Harsono}, {Hull}, {Hogerheijde}, {Murillo},
  \& {Tobin}}]{Tychoniec2021A&A...655A..65T}
{Tychoniec}, {\L}., {van Dishoeck}, E.~F., {van't Hoff}, M. L.~R., {et~al.}
  2021, \aap, 655, A65, \dodoi{10.1051/0004-6361/202140692}

\bibitem[{{Urquhart} {et~al.}(2018){Urquhart}, {K{\"o}nig}, {Giannetti},
  {Leurini}, {Moore}, {Eden}, {Pillai}, {Thompson}, {Braiding}, {Burton},
  {Csengeri}, {Dempsey}, {Figura}, {Froebrich}, {Menten}, {Schuller}, {Smith},
  \& {Wyrowski}}]{Urquhart2018MNRAS.473.1059U}
{Urquhart}, J.~S., {K{\"o}nig}, C., {Giannetti}, A., {et~al.} 2018, \mnras,
  473, 1059, \dodoi{10.1093/mnras/stx2258}

\bibitem[{{van der Tak} {et~al.}(2003){van der Tak}, {Boonman}, {Braakman}, \&
  {van Dishoeck}}]{van2003A&A...412..133V}
{van der Tak}, F.~F.~S., {Boonman}, A.~M.~S., {Braakman}, R., \& {van
  Dishoeck}, E.~F. 2003, \aap, 412, 133, \dodoi{10.1051/0004-6361:20031409}

\bibitem[{{van der Tak} {et~al.}(2020){van der Tak}, {Lique}, {Faure}, {Black},
  \& {van Dishoeck}}]{van2020Atoms...8...15V}
{van der Tak}, F. F.~S., {Lique}, F., {Faure}, A., {Black}, J.~H., \& {van
  Dishoeck}, E.~F. 2020, Atoms, 8, 15, \dodoi{10.3390/atoms8020015}

\bibitem[{{van Dishoeck} \& {Blake}(1998)}]{vanDi1998ARA&A..36..317V}
{van Dishoeck}, E.~F., \& {Blake}, G.~A. 1998, \araa, 36, 317,
  \dodoi{10.1146/annurev.astro.36.1.317}

\bibitem[{{van Gelder} {et~al.}(2021){van Gelder}, {Tabone}, {van Dishoeck}, \&
  {Godard}}]{vanGelder2021A&A...653A.159V}
{van Gelder}, M.~L., {Tabone}, B., {van Dishoeck}, E.~F., \& {Godard}, B. 2021,
  \aap, 653, A159, \dodoi{10.1051/0004-6361/202141591}

\bibitem[{{Vastel} {et~al.}(2018){Vastel}, {Qu{\'e}nard}, {Le Gal}, {Wakelam},
  {Andrianasolo}, {Caselli}, {Vidal}, {Ceccarelli}, {Lefloch}, \&
  {Bachiller}}]{Vastel2018MNRAS.478.5514V}
{Vastel}, C., {Qu{\'e}nard}, D., {Le Gal}, R., {et~al.} 2018, \mnras, 478,
  5514, \dodoi{10.1093/mnras/sty1336}

\bibitem[{{Vidal} {et~al.}(2017){Vidal}, {Loison}, {Jaziri}, {Ruaud},
  {Gratier}, \& {Wakelam}}]{Vidal2017MNRAS.469..435V}
{Vidal}, T. H.~G., {Loison}, J.-C., {Jaziri}, A.~Y., {et~al.} 2017, \mnras,
  469, 435, \dodoi{10.1093/mnras/stx828}

\bibitem[{{Vidal} \& {Wakelam}(2018)}]{Vidal2018MNRAS.474.5575V}
{Vidal}, T. H.~G., \& {Wakelam}, V. 2018, \mnras, 474, 5575,
  \dodoi{10.1093/mnras/stx3113}

\bibitem[{{Viti} {et~al.}(2001){Viti}, {Caselli}, {Hartquist}, \&
  {Williams}}]{Viti2001A&A...370.1017V}
{Viti}, S., {Caselli}, P., {Hartquist}, T.~W., \& {Williams}, D.~A. 2001, \aap,
  370, 1017, \dodoi{10.1051/0004-6361:20010300}

\bibitem[{{Wakelam} {et~al.}(2004){Wakelam}, {Caselli}, {Ceccarelli}, {Herbst},
  \& {Castets}}]{Wakelam2004A&A...422..159W}
{Wakelam}, V., {Caselli}, P., {Ceccarelli}, C., {Herbst}, E., \& {Castets}, A.
  2004, \aap, 422, 159, \dodoi{10.1051/0004-6361:20047186}

\bibitem[{{Wakelam} {et~al.}(2005){Wakelam}, {Ceccarelli}, {Castets},
  {Lefloch}, {Loinard}, {Faure}, {Schneider}, \&
  {Benayoun}}]{Wakelam2005A&A...437..149W}
{Wakelam}, V., {Ceccarelli}, C., {Castets}, A., {et~al.} 2005, \aap, 437, 149,
  \dodoi{10.1051/0004-6361:20042566}

\bibitem[{{Wang} {et~al.}(2013){Wang}, {Bourke}, {Hogerheijde}, {van der Tak},
  {Benz}, {Megeath}, \& {Wilson}}]{Wang2013A&A...558A..69W}
{Wang}, K.~S., {Bourke}, T.~L., {Hogerheijde}, M.~R., {et~al.} 2013, \aap, 558,
  A69, \dodoi{10.1051/0004-6361/201322087}

\bibitem[{{Wilson} \& {Rood}(1994)}]{Wilson1994ARA&A..32..191W}
{Wilson}, T.~L., \& {Rood}, R. 1994, \araa, 32, 191,
  \dodoi{10.1146/annurev.aa.32.090194.001203}

\bibitem[{{Woods} {et~al.}(2015){Woods}, {Occhiogrosso}, {Viti},
  {Ka{\v{n}}uchov{\'a}}, {Palumbo}, \& {Price}}]{Woods2015MNRAS.450.1256W}
{Woods}, P.~M., {Occhiogrosso}, A., {Viti}, S., {et~al.} 2015, \mnras, 450,
  1256, \dodoi{10.1093/mnras/stv652}

\bibitem[{{Xu} {et~al.}(2024){Xu}, {Wang}, {Liu}, {Tang}, {Evans}, {Palau},
  {Morii}, {He}, {Sanhueza}, {Liu}, {Stutz}, {Zhang}, {Chen}, {Li},
  {G{\'o}mez}, {V{\'a}zquez-Semadeni}, {Li}, {Mai}, {Lu}, {Liu}, {Chen}, {Li},
  {Shi}, {Ren}, {Li}, {Garay}, {Bronfman}, {Dewangan}, {Juvela}, {Lee},
  {Zhang}, {Yue}, {Wang}, {Ge}, {Jiao}, {Luo}, {Zhou}, {Tatematsu}, {Chibueze},
  {Su}, {Sun}, {Ristorcelli}, \& {Toth}}]{Xu2024ApJS..270....9X}
{Xu}, F., {Wang}, K., {Liu}, T., {et~al.} 2024, \apjs, 270, 9,
  \dodoi{10.3847/1538-4365/acfee5}

\bibitem[{{Yamamoto}(2017)}]{Yamamoto2017iace.book.....Y}
{Yamamoto}, S. 2017, {Introduction to Astrochemistry: Chemical Evolution from
  Interstellar Clouds to Star and Planet Formation},
  \dodoi{10.1007/978-4-431-54171-4}

\bibitem[{{Yang} {et~al.}(2022){Yang}, {Green}, {Pontoppidan}, {Bergner},
  {Cleeves}, {Evans}, {Garrod}, {Jin}, {Kim}, {Kim}, {Lee}, {Sakai},
  {Shingledecker}, {Shope}, {Tobin}, \& {van
  Dishoeck}}]{Yang2022ApJ...941L..13Y}
{Yang}, Y.-L., {Green}, J.~D., {Pontoppidan}, K.~M., {et~al.} 2022, \apjl, 941,
  L13, \dodoi{10.3847/2041-8213/aca289}

\bibitem[{{Yue} {et~al.}(2021){Yue}, {Qin}, {Liu}, {Tang}, {Wu}, {Wang}, \&
  {Zhang}}]{Yue2021RAA....21...14Y}
{Yue}, Y.-H., {Qin}, S.-L., {Liu}, T., {et~al.} 2021, Research in Astronomy and
  Astrophysics, 21, 014, \dodoi{10.1088/1674-4527/21/1/14}

\bibitem[{{Zapata} {et~al.}(2019){Zapata}, {Garay}, {Palau}, {Rodr{\'\i}guez},
  {Fern{\'a}ndez-L{\'o}pez}, {Estalella}, \&
  {Guzm{\'a}n}}]{Zapata2019ApJ...872..176Z}
{Zapata}, L.~A., {Garay}, G., {Palau}, A., {et~al.} 2019, \apj, 872, 176,
  \dodoi{10.3847/1538-4357/aafedf}

\bibitem[{{Zeng} {et~al.}(2021){Zeng}, {Jim{\'e}nez-Serra}, {Rivilla},
  {Mart{\'\i}n-Pintado}, {Rodr{\'\i}guez-Almeida}, {Tercero}, {de Vicente},
  {Rico-Villas}, {Colzi}, {Mart{\'\i}n}, \&
  {Requena-Torres}}]{2021ApJ...920L..27Z}
{Zeng}, S., {Jim{\'e}nez-Serra}, I., {Rivilla}, V.~M., {et~al.} 2021, \apjl,
  920, L27, \dodoi{10.3847/2041-8213/ac2c7e}

\bibitem[{{Zhang} {et~al.}(2023){Zhang}, {Yang}, {Zhang}, {Cox}, {Zeng},
  {Murillo}, {Ohashi}, \& {Sakai}}]{Zhang2023ApJ...946..113Z}
{Zhang}, Z.~E., {Yang}, Y.-l., {Zhang}, Y., {et~al.} 2023, \apj, 946, 113,
  \dodoi{10.3847/1538-4357/acbdf7}

\end{thebibliography}
\bibliographystyle{aasjournal}

\end{document}